# The role of symmetries on the stability of power grids

Jingyang Fang[1], Feng Gao[1], Xu Yang[2], Stephan Goetz[3]

[1] School of Control Science and Engineering, Shandong University, Jinan, SD, China.

[2] School of Electrical Engineering, Xi'an Jiaotong University, Shaanxi, SN, China.

[3] Department of Engineering, University of Cambridge, Cambridge, CB, United Kingdom.

✉emails: jingyangfang@sdu.edu.cn, fgao@sdu.edu.cn, yangxu@mail.xjtu.edu.cn, smg84@cam.ac.uk

The stability of power grids principally depends on the synchronization of multiple generating units. So far, the relationship between grid stability and symmetries remains unclear. This article fills in the research gap by innovatively using the mathematical tool of group theory. We find that grid symmetries involve two aspects—network and component symmetries, which reflect the homogeneities of network structures and generating units, respectively. As disclosed, the stability of power grids has a tight bearing on network and component symmetries, where corresponding automorphism group and symmetric subgroup orders reflect their levels of symmetries, respectively. In the perspective of network symmetries, where grid stability and synchronization speed are dictated by the algebraic connectivity, the stability ranking list of grids with four generators is $\boldsymbol{K_4} > \boldsymbol{C_4} > \boldsymbol{K_{1,3}} > \boldsymbol{R_4} > \boldsymbol{P_4} > \boldsymbol{X_4}$, which fully agrees with that of network symmetries. In terms of component symmetries, the stability of power grids largely benefits from improved component symmetries yet with $\boldsymbol{R_4}$ as an exceptional case, where it is recommended to avoid the centralized effort of a single generator. These conclusions have significant implications for the control of modern power grids with distributed resources and controllable generators, including energy storage and renewable power generators interfaced by grid-following/-forming converters.

## I. Introduction

Broadly speaking, synchronization refers to the coordination of events, processes, or data to ensure consistency and proper order in a system where multiple components operate concurrently. Particularly, the synchronization of coupled oscillators in a network can be dated back to the research done by Huygens on *an odd kind of sympathy* between coupled pendulum clocks and keeps fascinating researchers till now[1]. This fascinating subject features multidisciplinary nature, covering natural and social sciences as well as various engineering applications. In natural sciences, coupled pendulum clocks, metronomes, and Josephson junctions in physics, chemical oscillations in chemistry, coupled cortical neurons, heart pacemaker cells, and flashing fireflies in biology are vivid examples[2–5]; As for social sciences, the mechanism of rhythmic applause and pedestrian crowd synchrony on London's Millennium bridge involves synchronization[6]; In engineering, the synchronization of coupled oscillators resides across almost all fields, including coupled circuit oscillators in radio technology and signal processing, vehicle coordination and robotic locomotion in control engineering, multiple generator and power converter synchronization in electrical engineering, and laser arrays in optical engineering[7–10].

Power grids are the most complex networks ever engineered by human being[11]. Despite with various types of instability, a power grid is largely said to be stable if all interconnected generating units remain in synchronism when subjected to large disturbances[11]. A loss of synchrony can trigger sustained oscillations, which may later be translated into cascading failures or even widespread blackouts without any prompt treatment[11, 12]. With the ongoing global effort of decarbonization, the trend of large-scale renewable energy integration continues[13]. Various power converters, which act as renewable energy generators, will stand in contrast to and coexist with conventional synchronous generators in the foreseeable future, resulting in the increased heterogeneity of generating units and parameters across the networks, thereby further complicating the conditions and mechanisms for a power grid to maintain synchronization[14, 15]. As such, a deeper understanding of synchronization is increasingly important in modern power grids.

The synchronization of generators comprises *Phase Synchronization* and *Frequency Synchronization* among others[2]. In the former case, all the phases of generators converge to a common value with identical frequencies; while the latter case only requires a constant frequency of each generator. On top of frequency synchronization, *Phase cohesiveness* further requires that the pairwise phase difference of every pair of generators is upper bounded by a certain constant, e.g., $\pi/2$ rad or 90°, which should converge in steady state yet not necessarily be zero[2]. Notably, phase cohesiveness is exactly what we expected in a stable power grid[16]. Under acceptable assumptions, an equivalence between power grids and non-uniform Kuramoto oscillators has been established, thus allowing stability analyses via a modification of the celebrated Kuramoto model[16, 17]. Through this approach, it has been derived that the phase transition from incoherency to synchrony occurs when the strength of the mutual coupling outweighs the difference of natural frequencies[16]. For a fairly simple grid with a complete interaction network and uniform transmission lines, this synchronous condition reduces to $K > K_{\text{critical}}$, where $K$ represents the coupling factor and $K_{\text{critical}}$ refers to a certain threshold value, as proven by Kuramoto[4]. However, embedding the assumptions of homogeneous components and a complete interaction network with identical tie lines into an analytical model of real-world power grids remains rarely valid, thereby calling for more in-depth research[18, 19].

The condition for the synchronization of generators, and hence the stability of power grids, depends on the topology and parameters of the network as well as the parameters and types of individual generators[16, 20]. Along this research direction, Dörfler *et al*. derive a unique, concise, and closed-form synchronization condition by use of algebraic graph theory and differential geometry, which improves upon the existing conditions advocated thus far and remains statistically correct for almost all networks[16]. From the perspective of generators, Motter *et al*. derive an optimized damping coefficient under the assumptions of partially homogenous generators with identical damping coefficients. They demonstrate that system stability can be enhanced by not only generator homogenization but also matching them with network structure and parameters, yet invalid under large disturbances[18]. While the contemporary dialog in research centers on the role of inertia, the more critical role of generator damping in achieving a stable state has been pointed out[15]. From the perspective of network structures, Menck *et al*. conclude that dead ends and trees can severally deteriorate grid stability[21]. Such stability analyses have important implications, as they set the basis for advanced control and optimization methods. For instance, the nodal frequency of generators and transmission lines can be optimized offline for better synchronization[22]. Through the changes of the coupling strength, network structure, and edge dynamics, it is even possible to dynamically identify and stabilize problematic generators at the expense of communications[23]. On top of that, the desired patterns of synchrony can be obtained via the modifications of coupling coefficients and natural frequencies[24].

According to the existing body of literature, it has been vaguely found that the heterogenous level of generators and networks has a bearing on synchronization. Specifically, the highly-symmetric balanced synchronization patterns, where generators are clustered with identical phases in each cluster and phases uniformly distributed around a unit circle in different clusters, has been identified and summarized[2]. Such symmetric balanced patterns are found to be important for harmonic synchronization and elimination. In addition, Pecora *et al*. find an inherent relationship between the cluster formation and network symmetry[25]. As a step further, the same authors analyze all possible cluster synchronization patterns[26]. However, the relevant research focus is on the types and numbers of clusters instead of stable conditions, where application scenarios are electro-optic networks rather than power grids. Another interesting conclusion is that the heterogeneity of generators can translate into stronger system stability, which is claimed to agree with converse symmetry breaking theory[27]. However, such a conclusion is drawn without mathematically investigating symmetries and stays valid when only subjected to small disturbances. Moreover, the presented case targets at system dynamic performances rather than stable conditions.

This article investigates the relationship between grid symmetries, including network and component symmetries, and the stability of power grids. The main contributions are twofold:

1) We innovatively apply the mathematical tool of group theory, as combined with graph theory, to quantitatively anaylyze the symmetries of power grids. Specifically, we investigate network and component symmetries via the automorphism groups of associated graphs and the subgroups of symmetric groups, respectively, where higher group orders imply improved symmetries;

2) Our analyses stay valid under not only small disturbances but also large disturbances. In terms of network symmetries, the stability ranking list of grids with four generators is $\boldsymbol{K_4} > \boldsymbol{C_4} > \boldsymbol{K_{1,3}} > \boldsymbol{R_4} > \boldsymbol{P_4} > \boldsymbol{X_4}$, which follows that of network symmetries. As for component symmetries, the stability of power grids largely benefits from improved component symmetries yet with an exception $\boldsymbol{R_4}$, which corresponds to the Klein Group and has a normal subgroup feature.

These conclusions, as verified by simulation results, have significant implications for the control of modern power grids with distributed resources and controllable generators, including energy storage and renewable power generators interfaced by grid-following/-forming converters.

## Results

## II. Power Grid Stability

### (1) Problem Statement

The stability of power grids mainly depends on the synchronization of all generators. Due to the inherent complexity of power grids, assumptions are always necessary for the simplification of stability analysis[28]. Under such assumptions, we can model generators as dynamic elements, while transmission lines are treated as algebraic elements[15]. As a result, a widely adopted structure-preserving model of power grids comprises a population of heterogeneous generators, a graph describing the interaction among them, and sinusoidal coupling, which takes the form of a modified Kuramoto model (see Methods for details) as[16, 28]

$$\frac{\mathrm{d}\theta_{\mathrm{gi}}(t)}{\mathrm{d}t} = \omega_{\mathrm{gi}} - \sum_{j=1}^{n} a_{\mathrm{ij}} \sin\left[\theta_{\mathrm{gi}}(t) - \theta_{\mathrm{gj}}(t)\right] \qquad i \in \{1,2,\dots,n\}\ , \tag{1}$$

where $\theta_{gi}(t)$ represents the phase angle of the $i$th generator, which falls into a range of [0, 2π) and repeats every 2π. $\omega_{gi} = p_{gi_ref}\omega_{g0}/(S_{gi}D_{gi})$ represents the natural angular frequency of the $i$th generator, where $p_{gi_ref}$ stands for the input active power reference, $D_{gi}$ represents the damping coefficient, $\omega_{g0}$ refers to the nominal angular frequency, and $S_{gi}$ designates the rated power. $a_{ij} = 3v_{gi}v_{gj}\omega_{g0}/(2S_{gi}X_{gij}D_{gi})$ models the coupling strength between the $i$th generator and the $j$th generator, where $v_{gi}$ and $v_{gj}$ designate the root-mean-square (RMS) terminal voltages of generators, and $X_{gij}$ refers to the line reactance between the two generators. Note that Eq. (1) applies equally well to coupled synchronous generators and grid-forming converters[29].

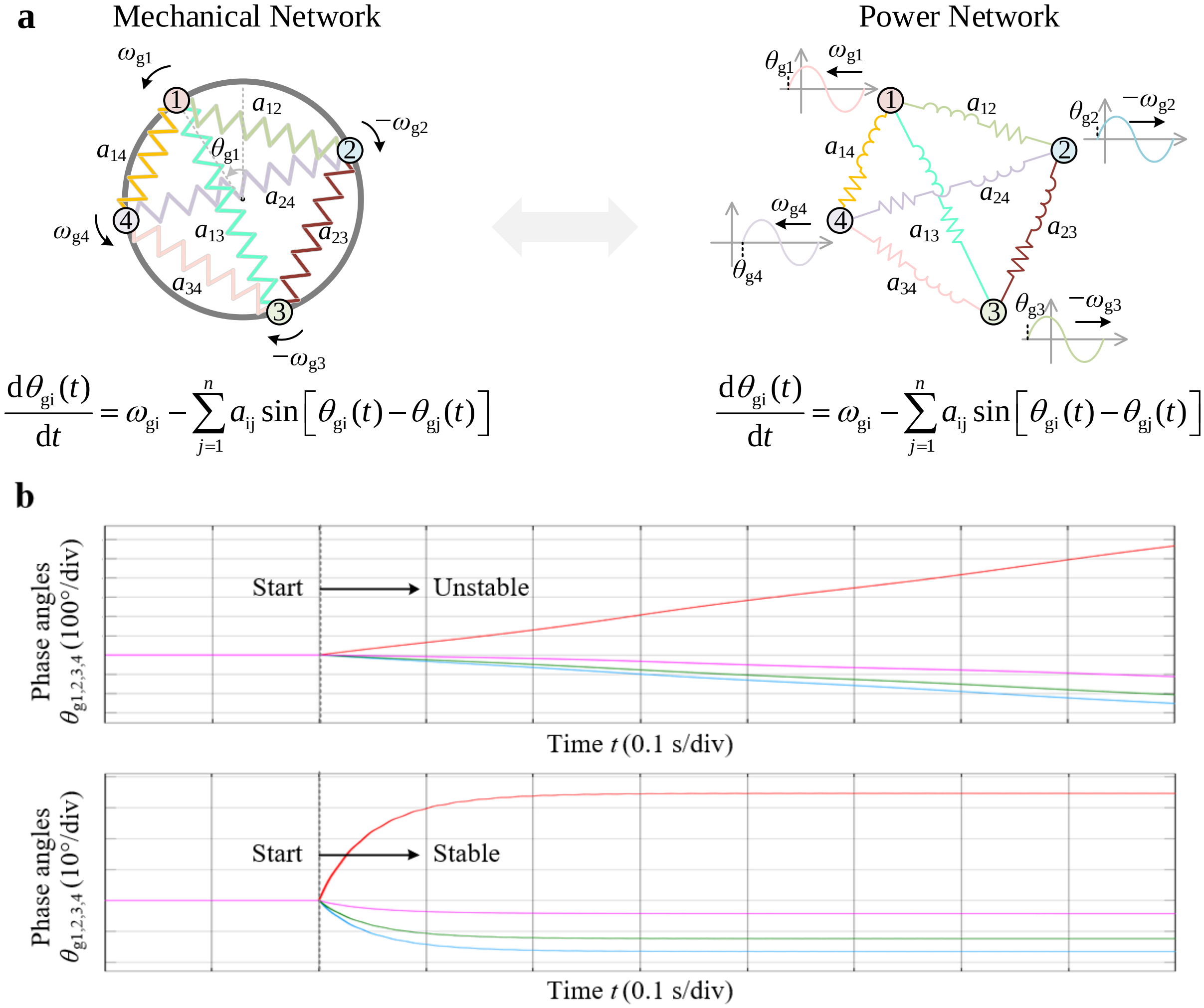

**Fig. 1 An analogy between the mechanical network with four coupled oscillating particles and the power network with four coupled generators as well as their typical phase angle responses in the presence of a large disturbance**. **a** The schematic diagrams of the mechanical network with four particles (in the left-hand side), where the network consists of four oscillating particles constrained to rotate around a circle without colliding, and the power network with four generators (in the right-hand side), which are interconnected via transmission lines. The dynamics of both systems can mathematically be described by the same modified Kuramoto model Eq. (1), where $\theta_{gi}$ represents the phase angle of the *i*th particle/generator, $\omega_{gi}$ represents the natural angular frequency of the *i*th particle/generator, $a_{ij}$ stands for the coupling strength of the spring/transmission line connecting the *i*th particle/generator and the *j*th particle/generator. **b** Two typical phase angle response plots of the mechanical/power network in the presence of a large disturbance. The upper plot shows an unstable case with weakly coupled oscillators and strongly heterogeneous natural frequencies, leading to divergent phase angle differences; The lower plot illustrates a stable case with strongly coupled oscillators and sufficiently homogeneous natural frequencies, leading to convergent phase angle differences, which are confined within 90 degrees during transients.

Fig. 1a visualizes an analogy between the schematic diagrams of the mechanical network with coupled oscillating particles and the power network with coupled generators, both of which can mathematically be described by Eq. (1). As for the mechanical network, four oscillating particles are constrained to rotate around a unit circle without colliding and interconnected via springs[2]. These particles are subject to a competition between the natural frequency exerted by the external force $\omega_{gi}$ and the spring coupling $a_{ij}\sin[\theta_{gi}(t) - \theta_{gj}(t)]$ contributed by the internal restoring force, leading to a tradeoff between each particle's tendency to follow its natural frequency and the synchronization effort of its neighbors[16]. Despite with different physical meanings of parameters, the model, contributing factors, and solutions apply equally well to the power grid with four generators coupled by transmission lines, thus implying a mapping between the mechanical network and power network.

Fig. 1b demonstrates two typical phase angle response plots of the mechanical/power network in the presence of a large disturbance. The upper plot shows an unstable case with weakly coupled oscillators and strongly heterogeneous natural frequencies, leading to divergent phase angles, which translates into instability in practice. Alternatively, the lower plot represents a stable case with strongly coupled oscillators and sufficiently homogeneous natural frequencies, leading to a unified frequency in steady state and convergent phase angle differences, which are confined within 90° during transients, thereby representing a stable system. In both cases, the dynamics of phase angles are determined by the modified Kuramoto model of Eq. (1).

**(2) Stable Condition**

As a natural next step, we will present the stable condition of power grids. For a stable power grid, all generators should oscillate with their synchronized frequencies being identical, namely,

$$\dot{\theta}_{\mathrm{gi}}(t) = \omega_{\mathrm{syn}} = \frac{1}{n}\sum_{j=1}^{n}\omega_{\mathrm{gj}}, \forall i \in \{1, 2, ..., n\}, \tag{2}$$

where $\omega_{\mathrm{syn}}$ stands for the synchronized angular frequency. In this case, we can substitute $\theta_{\mathrm{gi}}(t)$ by $\theta_{\mathrm{gi}}(t)-\omega_{\mathrm{syn}}t$ in Eq. (1) and reorganize it in a rotating reference frame, which features a zero synchronized frequency. Therefore, without loss of generality, we will assume $\omega_{\mathrm{syn}} = 0$ from now on. On top of synchronized frequencies, the phase angles of generators in a stable power grid are required to be cohesive, i.e., $|\theta_{\mathrm{gi}}(t)-\theta_{\mathrm{gj}}(t)| \leq \gamma < \pi/2$, where the phase angle difference for every pair of the interconnected generators $i$ and $j$ is upper bounded by $\pi/2$[2]. Otherwise, the control relationship between active power and phase angle reverses, leading to the malfunction of frequency regulator and causing instablity[11].

By use of algebraic graph theory, we can mathematically describe the interconnection topology and coupling strength of a power grid by a connected, undirected, and weighted graph $\mathbf{G}(\mathbf{V}, \mathbf{E})$, where $\mathbf{V} = \{1, 2, \ldots, n\}$ stands for the set of nodes, and $\mathbf{E} \subset \mathbf{V}\times\mathbf{V}$ contains the interconnected node pairs and represents the set of edges[30]. Note that the edge between two generators exists, i.e., $\{i, j\} \in \mathbf{E}$, if and only if they are directly connected via a transmission line. The adjacency matrix $\mathbf{A}$ with elements $a_{\mathrm{ij}}$ describes the interconnection relationship of a graph, and the Laplacian matrix $\mathbf{L}$ evaluates the difference between each node and its neighbors, whose detailed expression can be found in Methods.

According to Dörfler *et al*., the power grid modelled by Eq. (1) has a unique and stable solution $\boldsymbol{\theta}_{\mathbf{g}}^* = [\theta_{\mathrm{g1}}^*\ \theta_{\mathrm{g2}}^* \ldots \theta_{\mathrm{gn}}^*]^{\mathrm{T}}$ with synchronized frequencies and cohesive phases if[16]

$$\left\|\mathbf{L}^{\dagger}\boldsymbol{\omega}_{\mathbf{g}}\right\|_{\mathbf{E},\infty} \leq \sin\gamma, \tag{3}$$

where $\mathbf{L}^{\dagger}$ stands for the pseudoinverse matrix of $\mathbf{L}$, which satisfies $\mathbf{L}\mathbf{L}^{\dagger}\mathbf{L} = \mathbf{L}$ and $\mathbf{L}^{\dagger}\mathbf{L}\mathbf{L}^{\dagger} = \mathbf{L}^{\dagger}$. $\boldsymbol{\omega}_{\mathbf{g}} = [\omega_{\mathrm{g1}}\ \omega_{\mathrm{g2}} \ldots \omega_{\mathrm{gn}}]^{\mathrm{T}}$ represents the vector of natural angular frequencies. $\gamma$ denotes the maximum allowable phase angle distance. $\|\mathbf{x}\|_{\mathbf{E},\infty} = \max_{\{i,j\}\in\mathbf{E}}|x_{\mathrm{i}} - x_{\mathrm{j}}|$ is the worst-case dissimilarity for the vector $\mathbf{x} = [x_1\ x_2 \ldots x_{\mathrm{n}}]^{\mathrm{T}}$ over the edges $\{i, j\} \in \mathbf{E}$.

Under the assumption of DC power flow, where sinusoidal coupling terms in Eq. (1) $a_{\mathrm{ij}}\sin[\theta_{\mathrm{gi}}(t) - \theta_{\mathrm{gj}}(t)]$ are linearized by $a_{\mathrm{ij}}[\theta_{\mathrm{gi}}(t) - \theta_{\mathrm{gj}}(t)]$, it is straightforward to derive Eq. (3) from Eq. (1), with the latter expressed in vector notation as $\mathrm{d}\boldsymbol{\theta}_{\mathbf{g}}(t)/\mathrm{dt} = \boldsymbol{\omega}_{\mathbf{g}} - \mathbf{L}\boldsymbol{\theta}_{\mathbf{g}}(t)$, yielding a steady state solution $\boldsymbol{\theta}_{\mathbf{g}}^* = \mathbf{L}^{\dagger}\boldsymbol{\omega}_{\mathbf{g}}$. Despite this clear linearized interpretation, note that Eq. (3) stays valid under large disturbances[16].

When $\gamma = \pi/2$, which holds in most cases, the right-hand-side term of Eq. (3) becomes $\sin(\gamma) = 1$. Under this scenario, the stability of power grids solely depends on $\mathbf{L}$ and $\boldsymbol{\omega}_\mathbf{g}$. Remarkably, the network structure and parameters determine the Laplacian matrix $\mathbf{L}$, while the types and parameters of power grid components, including conventional synchronous and renewable generators, affect the vector of natural angular frequencies $\boldsymbol{\omega}_\mathbf{g}$. As claimed by Dörfler *et al.*, the stable condition in Eq. (3) remains as the sharpest solution[2, 16]. For a classic Kuramoto oscillator model with a complete graph and uniform gains, where $a_{ij} = K/n$, the stable condition of Eq. (3) reduces to $K > \max_{\{i,\, j\}\in\{1,\, 2,\, \ldots,\, n\}}|\omega_{gi} - \omega_{gj}|$, which agrees well with the results proven by Kuramoto[4]. A sufficient condition for Eq. (3) to be true is the algebraic connectivity condition[16]:

$$\lambda_2(\mathbf{L}) \geq \left\|\boldsymbol{\omega}_\mathbf{g}\right\|_{\mathbf{E},\infty}, \tag{4}$$

where $\lambda_2(\mathbf{L})$, also known as the algebraic connectivity, denotes the second smallest eigenvalue of $\mathbf{L}$. Understandably, network structure and parameters determine $\lambda_2(\mathbf{L})$. In parallel, $||\boldsymbol{\omega}_\mathbf{g}||_{\mathbf{E},\infty}$ is mainly affected by the types and parameters of grid components. It is clear from Eq. (4) that a larger $\lambda_2(\mathbf{L})$ and a smaller $||\boldsymbol{\omega}_\mathbf{g}||_{\mathbf{E},\infty}$ translates into better stability. As such, these two parts will be investigated in detail in later sections.

## III. Group Theory and Automorphism Groups

We aim to disclose the relationship between the stability and symmetries of power grids. To achieve this, the background knowledge of group theory is necessary, which serves as a fundamental mathematical tool to study symmetries.

### (1) Fundamentals of Group Theory

Loosely speaking, symmetries refer to certain operations that maintain objects unchanged. For visualization, Fig. 2 shows typical symmetric operations along with their corresponding natural and artificial symmetry examples, including the mirror symmetry, rotational symmetry, translational symmetry, and scaling symmetry. Specifically, the mirror symmetry flips the object, the rotational symmetry rotates the object by a certain degree, the translational symmetry moves the object a distance along a selected direction, and the scaling symmetry zooms in or out the object, where the object remains unchanged after each operation.

Group theory serves as the key mathematical tool to investigate symmetries. By definition, a **group** refers to a set $G$ together with a binary multiplication operation $\circ$, which possesses the following properties[31]:

1) $\forall a,b \in G$, $a \circ b \in G$ (closure);

2) $\forall a,b,c \in G$, $a \circ b \circ c = a \circ (b \circ c)$ (associativity);

3) $\forall a \in G$, $\exists e \in G$ such that $a \circ e = e \circ a = a$ (existence of identity);

4) $\forall a \in G$, $\exists a^{-1} \in G$ such that $a \circ a^{-1} = a^{-1} \circ a = e$ (existence of inverses);

One notable group is the **symmetric group** $S_n$, which contains all the symmetries related to the function composite operation that permutates over the set $\{1, 2, \ldots, n\}$. For example, $S_3$ includes six elements

$$\left\{ s_1 = \begin{pmatrix} 1 & 2 & 3 \\ 1 & 2 & 3 \end{pmatrix},\ s_2 = \begin{pmatrix} 1 & 2 & 3 \\ 2 & 3 & 1 \end{pmatrix},\ s_3 = \begin{pmatrix} 1 & 2 & 3 \\ 3 & 1 & 2 \end{pmatrix},\ s_4 = \begin{pmatrix} 1 & 2 & 3 \\ 1 & 3 & 2 \end{pmatrix},\ s_5 = \begin{pmatrix} 1 & 2 & 3 \\ 3 & 2 & 1 \end{pmatrix},\ s_6 = \begin{pmatrix} 1 & 2 & 3 \\ 2 & 1 & 3 \end{pmatrix} \right\}.$$

For this two-line notation, the first line refers to the sequence before operation, while the second line refers to the sequence after operation. Let us take the second element $s_2$, which corresponds to the function $\sigma$ that satisfies $\sigma(1) = 2$, $\sigma(2) = 3$, and $\sigma(3) = 1$. Similar to the composition operation of functions, the multiplication operation of group elements (from right to left) is defined accordingly. Suppose that the function associated with $s_3$ is $\tau$. We have $s_2 \circ s_3 = s_1$, because $\sigma(\tau(1)) = \sigma(3) = 1$, $\sigma(\tau(2)) = \sigma(1) = 2$, and $\sigma(\tau(3)) = \sigma(2) = 3$. Building on the group multiplication operation, we tabulate Table I that describes the multiplication operation of every pair of group elements. Referring to Table I, we can easily verify the closure property, the associativity property, the existence of identity $e = s_1$, and the existence of inverses, thereby confirming that $S_3$ is indeed a group. Note that the **group order** is defined as the number of elements in a group. In the case of $S_n$, its order is $|S_n| = n!$. In particular, $|S_3| = 6$.

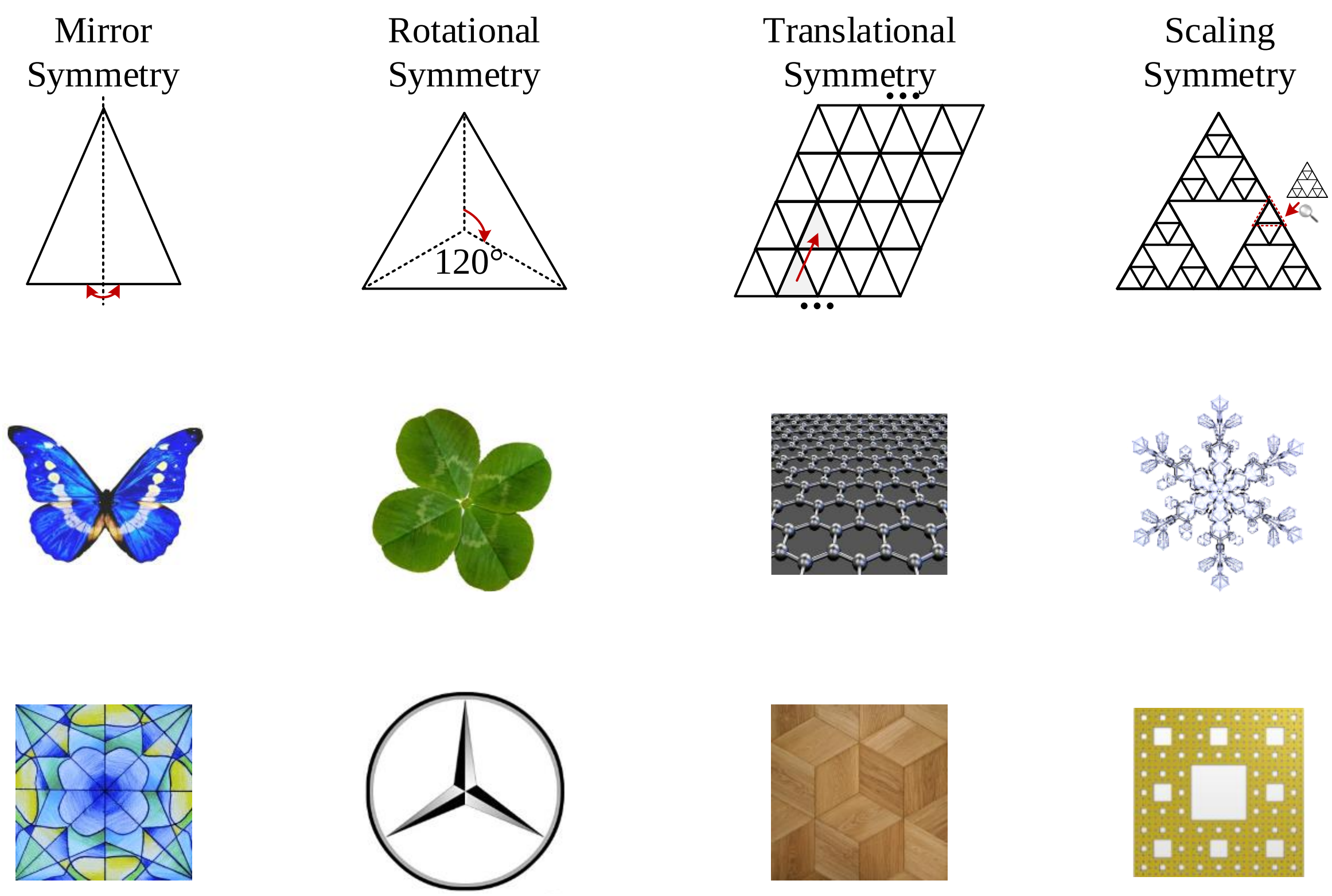

**Fig. 2 Typical symmetries, including the mirror symmetry, rotational symmety, translational symmetry, and scaling symmetry**. The figures in the first row visualizes symmetric operations, where the mirror symmetry flips the object, the rotational symmetry rotates the object by a certain degree (e.g., 120 degrees), the translational symmetry moves the object a distance along a selected direction, and the scaling symmetry zooms in or out the object, which remains unchanged after each operation. The figures in the second row presents four examples of natural symmetries, including (from left to right) a butterfly with the mirror symmetry, a clover with the rotational symmetry, the graphene with the translational symmetry, and a snowflake with the scaling symmetry. The figures in the third row shows four examples of artificial symmetries, including (from left to right) the stained glass with the mirror symmetry, the Benz vehicle sign with the rotational symmetry, floorboards with the translational symmetry, and a printed-circuit board with the scaling symmetry.

**Table I Cayley Table of the 3-order symmetric group $S_3$.** The first column contains the first element of the group multiplication. The first row contains the second element of the group multiplication**.** The remaining elements correspond to the results of group multiplications.

| Column $\circ$ Row | $s_1$ | $s_2$ | $s_3$ | $s_4$ | $s_5$ | $s_6$ |
|---|---|---|---|---|---|---|
| $s_1$ | $s_1$ | $s_2$ | $s_3$ | $s_4$ | $s_5$ | $s_6$ |
| $s_2$ | $s_2$ | $s_3$ | $s_1$ | $s_6$ | $s_4$ | $s_5$ |
| $s_3$ | $s_3$ | $s_1$ | $s_2$ | $s_5$ | $s_6$ | $s_4$ |
| $s_4$ | $s_4$ | $s_5$ | $s_6$ | $s_1$ | $s_2$ | $s_3$ |
| $s_5$ | $s_5$ | $s_6$ | $s_4$ | $s_3$ | $s_1$ | $s_2$ |
| $s_6$ | $s_6$ | $s_4$ | $s_5$ | $s_2$ | $s_3$ | $s_1$ |

Next, we introduce another important concept—**subgroup**. A subgroup $H$ is a group that inherits some elements (including the identity) of $G$ together with its binary operation $\circ$, denoted as $H \leq G$. Clearly, the identity element $e$ and the group $G$ itself are two trivial subgroups, while other subgroups are nontrivial subgroups. As a representative subgroup of $S_n$, the **cyclic group** $C_n$ is formed by a single generator and contains only the cyclic rotational elements. For instance, we select a generator $s_2$ from $S_3$, together with the identity element $e$. According to the closure property, the multiplication of $s_2$ by itself, i.e., $s_2 \circ s_2 = s_3$, must be included to form a group, leading to the 3rd-order cyclic group $C_3$, which includes three elements $\left\{ c_1 = s_1 = s_2 \circ s_2^{-1} = \begin{pmatrix} 1 & 2 & 3 \\ 1 & 2 & 3 \end{pmatrix}, c_2 = s_2 = \begin{pmatrix} 1 & 2 & 3 \\ 2 & 3 & 1 \end{pmatrix}, c_3 = s_3 = s_2 \circ s_2 = \begin{pmatrix} 1 & 2 & 3 \\ 3 & 1 & 2 \end{pmatrix} \right\}$.

The Cayley Table of the 3-order cyclic group $C_3$ covers the part related to $s_1$, $s_2$, and $s_3$ in Table I. In consequence, it is straightforward to verify all the four group properties of $C_3$. In general, $|C_n| = n$. According to the **Lagrange's Theorem in Group Theory**, the orders of all subgroups of $G$ must be divided by the order of $G$[31]. In other words, $\forall\, H \leq G$, $|G| = k|H|$, where $k$ is a positive integer.

**(2) Isomorphism and Automorphism Groups**

To further investigate the symmetry of network graphs, automorphism groups serve as a useful tool.

Before presenting automorphism groups, we should be familiar with the concept of isomorphism. With two groups $G_1$ and $G_2$, define $f : G_1 \rightarrow G_2$ as a function that maps each element of $G_1$ to a specific element in $G_2$. If $\forall a, b \in G_1$, $f(a \circ b) = f(a) \circ f(b)$. Then, $f$ is a **homomorphism** from $G_1$ to $G_2$. Furthermore, **isomorphism** refers to a bidirectional homomorphism with a one-to-one function $f : G_1 \rightarrow G_2$ and $f^{-1} : G_2 \rightarrow G_1$, denoted as $G_1 \cong G_2$.

Next, we transplant the concept of isomorphism into graphs. For two simple graphs $\mathbf{G_1}(\mathbf{V_1}, \mathbf{E_1})$ and $\mathbf{G_2}(\mathbf{V_2}, \mathbf{E_2})$, if there exists a one-to-one function $f : \mathbf{V_1} \rightarrow \mathbf{V_2}$ and $f^{-1} : \mathbf{V_2} \rightarrow \mathbf{V_1}$, which satisfies $\forall a, b \in \mathbf{V_1}$, $(f(a), f(b)) \in \mathbf{E_2}$ if and only if $(a, b) \in \mathbf{E_1}$, the two graphs $\mathbf{G_1}$ and $\mathbf{G_2}$ are isomorphic[30]. When graphs are isomorphic, their vertices keep the one-to-one adjacent relationship. Therefore, isomorphism implies the identical numbers of nodes, edges, and node degrees of two graphs. Formally, an **automorphism** of a graph $\mathbf{G}(\mathbf{V}, \mathbf{E})$ is a permutation $\sigma$ of the vertex set $\mathbf{V}$, such that the pair of vertices forms an edge $(a, b) \in \mathbf{E}$ if and only if $(\sigma(a), \sigma(b)) \in \mathbf{E}$. Automorphism targets at one graph, which is isomorphic to itself after a vertex permutation. Therefore, two vertices are adjacent if and only if their images after permutation are adjacent.

Finally, the **automorphism group** $\mathrm{Aut}(\mathbf{G})$ of a graph $\mathbf{G}$ refers to the group that is formed by all the vertex permutations that preserve the edge-vertex connectivity. Mathematically speaking, for a permutation $\sigma$ that satisfies $(a, b) \in \mathbf{E}$ if and only if $(\sigma(a), \sigma(b)) \in \mathbf{E}$, then $\sigma \in \mathrm{Aut}(\mathbf{G})$. Importantly, the degree of $\mathrm{Aut}(\mathbf{G})$ can reflect the level of symmetries for a graph, which lays foundation for the subsequent analysis of graph symmetries. According to **Cayley's Theorem**, every group $G$ is isomorphic to a subgroup of the symmetric group $S_n$. Therefore, on the one end of the spectrum, a complete graph $\boldsymbol{K_n}$ possesses the highest level of symmetries, as $|\mathrm{Aut}(\boldsymbol{K_n})| = |S_n| = n!$. On the other end of the spectrum, the asymmetric graph $\boldsymbol{X_n}$ only contains the trivial automorphism without any change of vertices, leading to $|\mathrm{Aut}(\boldsymbol{X_n})| = 1$.

## IV. Symmetries and Stability

In this section, we will investigate the role of symmetries on grid stability by use of the tools discussed before. First, we will explore network symmetries. Second, we will investigate component symmetries. Third, we will analyze the relationship between symmetries and stability.

### (1) Network Symmetries

Recapping Fig. 1, we choose a simplified power grid comprising four generators and their interconnected transmission lines as an example. Fig. 3 illustrates all possible power network graphs with different symmetries, including the complete graph $\boldsymbol{K_4}$, the cyclic graph $\boldsymbol{C_4}$, the star graph $\boldsymbol{K_{1,3}}$, the rectangular graph $\boldsymbol{R_4}$, the path graph $\boldsymbol{P_4}$, and the random graph $\boldsymbol{X_4}$, and the associated Cayley graphs. Among these graphs, the complete graph $\boldsymbol{K_4}$ features the identical edges linking all pairs of vertices. As mentioned, its automorphism group is a symmetric group of order 4 with $|S_4| = 24$, as all vertex permutations preserve the edge-vertex connectivity. In consequence, the complete graph $\boldsymbol{K_4}$ possesses the highest level of symmetries. The other graphs have fewer symmetries, which correspond to the subgroups of $S_4$. Specifically, the cyclic graph $\boldsymbol{C_4}$ comprises a loop connecting individual vertices, and its automorphism group is the dihedral group of order 4 with $|D_4| = 8$. The star graph $\boldsymbol{K_{1,3}}$ possesses a central vertex connected by the other vertices. Its automorphism group is the symmetric group of order 3 with $|S_3| = 6$, as the permutations of three identical vertices preserve the adjacency. The rectangular graph $\boldsymbol{R_4}$ comprises two different pairs of symmetrical edges and vertices, whose automorphism group is the Klein group of order 4 with $|V_4| = 4$. The path graph $\boldsymbol{P_4}$ has two different pairs of symmetrical vertices, whose automorphism group is the cyclic group of order 2 with $|C_2| = 2$, which possesses only one reflection symmetry. Evidently, no symmetry can be identified in the random graph $\boldsymbol{X_4}$, whose automorphism group is the trivial group with $|C_1| = 1$. In Fig. 3, the Cayley graphs shown in the right-hand side serve as a visual aid and encode the abstract structure of groups. For instance, the dihedral group $D_4$, as the automorphism group of the cyclic graph $\boldsymbol{C_4}$, feature 2 rotational symmetries with order 4 in clockwise and counter-clockwise directions, thereby satisfying $|D_4| = 8$.

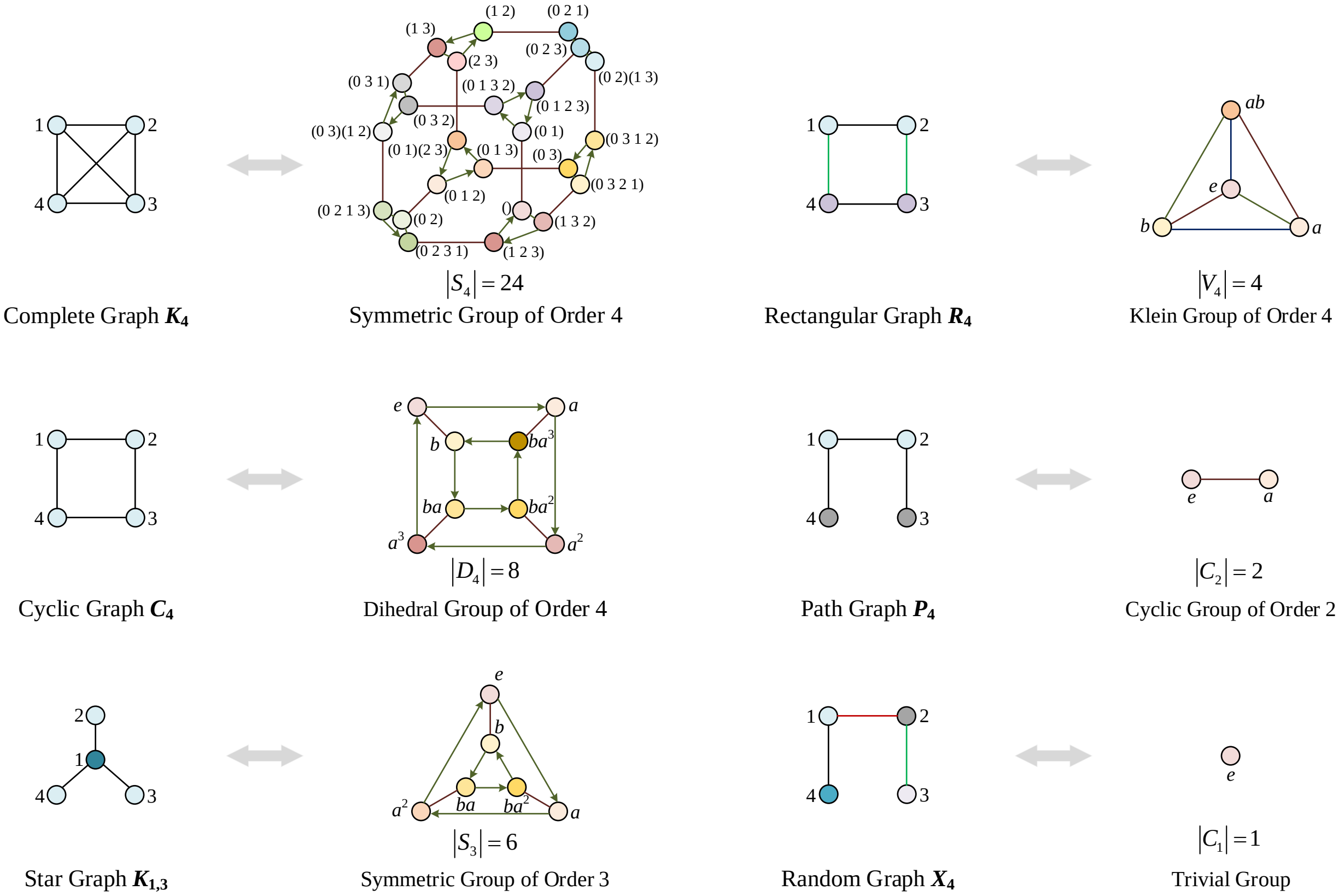

**Fig. 3 Power network graphs with different symmetries, including the complete graph $K_4$, cyclic graph $C_4$, star graph $K_{1,3}$, rectangular graph $R_4$, path graph $P_4$, and random graph $X_4$, and associated Cayley graphs as a visual aid**. Among these graphs, the complete graph $\boldsymbol{K_4}$ features the identical edges linking all pairs of vertices. Its automorphism group is the symmetric group of order 4 with $|S_4| = 24$ and possesses the highest level of symmetries. The cyclic graph $\boldsymbol{C_4}$ features a loop connecting individual vertices. Its automorphism group is the dihedral group of order 4 with $|D_4| = 8$. The star graph $\boldsymbol{K_{1,3}}$ possesses a central vertex connected by the other vertices. Its automorphism group is the symmetric group of order 3 with $|S_3| = 6$. The rectangular graph $\boldsymbol{R_4}$ comprises two different pairs of edges and vertices, whose automorphism group is the Klein group of order 4 with $|V_4| = 4$. The path graph $\boldsymbol{P_4}$ has two different pairs of vertices. Its automorphism group is the cyclic group of order 2 with $|C_2| = 2$. The random graph $\boldsymbol{R_4}$ has no symmetry. Its automorphism group is the trivial group with $|C_1| = 1$.

Based on the graphs with different symmetries shown in Fig. 3, we continue to derive the corresponding algebraic connectivity $\lambda_2$ of the relevant Laplacian matrix $\mathbf{L}$ and the pseudoinverse Laplacian matrices $\mathbf{L}^{\dagger}$, which are necessary for stability evaluation as per Eq. (4) and Eq. (3), respectively. The matrices associated with all graphs in Fig. 3 are derived and listed in Methods. Notably, for a fair comparison of symmetries, we unify the value of edges, thereby allowing the disclosure of the inherent relationship between network symmetries and grid stability.

By definition, $\lambda_2$ refers to the second smallest eigenvalue (counting multiple eigenvalues separately) of $\mathbf{L}$. As verified in Methods, the ranking list of network graphs in terms of the algebraic connectivity is $\lambda_2(\boldsymbol{K_4}) > \lambda_2(\boldsymbol{C_4}) > \lambda_2(\boldsymbol{K_{1,3}}) > \lambda_2(\boldsymbol{R_4}) > \lambda_2(\boldsymbol{P_4}) > \lambda_2(\boldsymbol{X_4})$, which totally agrees with the ranking list of symmetries. Notably, the greater the algebraic connectivity, the faster and more stable the synchronization of generators[30].

Fig. 4 presents the heatmaps of the per-unit adjacency matrix $\mathbf{A}$, Laplacian matrix $\mathbf{L}$, and pseudoinverse Laplacian matrices $\mathbf{Li}$ (i.e., $\mathbf{L}^{\dagger}$) of the power network graphs with different symmetries, including the complete graph $\boldsymbol{K_4}$, cyclic graph $\boldsymbol{C_4}$, star graph $\boldsymbol{K_{1,3}}$, rectangular graph $\boldsymbol{R_4}$, path graph $\boldsymbol{P_4}$, and random graph $\boldsymbol{X_4}$. Recapping Eq. (3), we find that the pseudoinverse Laplacian matrix $\mathbf{L}^{\dagger}$ plays a decisive role on the stability of power grids, which also depends on the vector of natural angular frequencies $\boldsymbol{\omega}_\mathbf{g}$, as determined by the types and parameters of generators and will be analyzed in the following part. In Fig. 4, it is roughly true that the darker the colour of **Li**, the higher chance of instability. From this perspective, the ranking list of network graphs largely agrees with the ranking list of network graphs in terms of symmetries. This implies the inherent positive relationship between network symmetries and stability of power grids. However, note that Eq. (3) and Eq. (4) are sufficient conditions and thus involve conservativeness.

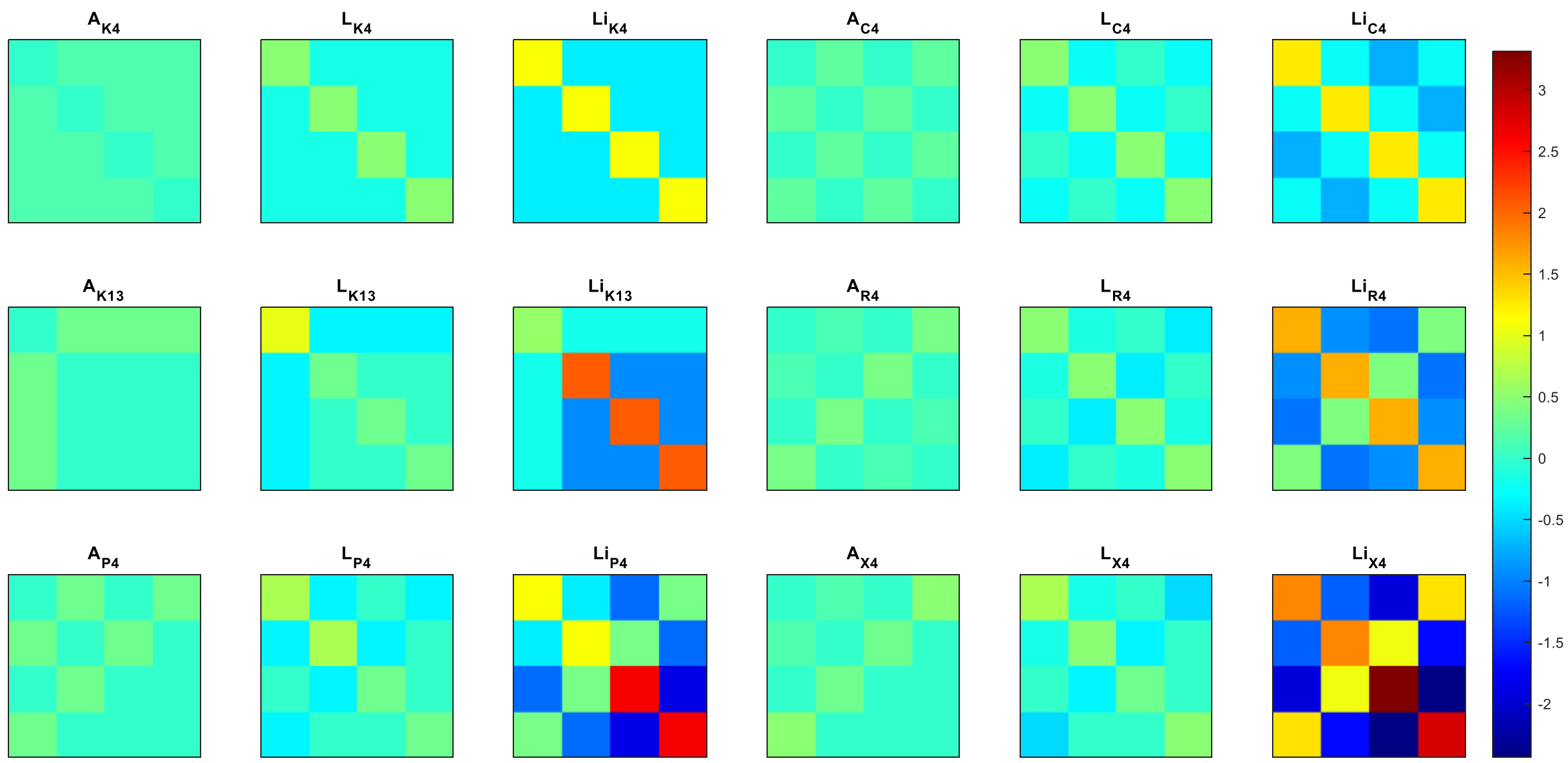

**Fig. 4 Heatmaps of the adjacency matrix A, Laplacian matrix L, and pseudoinverse Laplacian matrix Li of the power network graphs with different symmetries, including the complete graph $K_4$, cyclic graph $C_4$, star graph $K_{1,3}$, rectangular graph $R_4$, path graph $P_4$, and random graph $X_4$.** The adjacency matrix **A** describes the interconnection relationships between every pair of generators. The Laplacian matrix **L** can be derived by subtraction of the relevant adjacency matrix **A** from the degree matrix. The pseudoinverse Laplacian matrix **Li** affects the stability of power grids. Loosely speaking, the darker the colour of **Li**, the higher chance of instability. From the perspective of the algebraic connectivity, the ranking list of network graphs should be $\boldsymbol{K_4} > \boldsymbol{C_4} > \boldsymbol{K_{1,3}} > \boldsymbol{R_4} > \boldsymbol{P_4} > \boldsymbol{X_4}$, which agrees with the ranking list in terms of symmetries.

**(2) Component Symmetries**

It is noticeable that the symmetries of generating units have a tight correlation with system stability. Obviously, in the extreme case where all generators are identical, they will always be synchronized and never face any transient instability problem, as their phase angles vary synchronously. In practice, generators suffer from inherent heterogeneities due to different generator types, operating conditions, and control parameters[15]. Such heterogeneities are exacerbated by the advancement of large-scale renewable integration, because renewable energies are normally connected to the grid through power converters instead of synchronous generators[33]. Besides, the aging disparities and retrofitting inconsistencies of generators further complicate the scenario. As a result, the ideal assumption of generator homogeneity is overly optimistic for real power grids.

According to Eq. (3) and Eq. (4), the heterogeneity of generators can be reflected by their vector of natural angular frequencies $\boldsymbol{\omega}_\mathbf{g}$. Once again, we employ the power grid with four generators as an example. Under this condition, $\boldsymbol{\omega}_\mathbf{g} = [\omega_{g1}\ \omega_{g2}\ \omega_{g3}\ \omega_{g4}]^T$. Here, we investigate its symmetries by the subgroups of $S_4$. According to the Lagrange's Theorem, the orders of all subgroups of $S_4$ must be divided by 24. However, it should be claimed that not all the subgroups of $S_4$ correspond to the symmetries of generators. In total, the symmetries of generators comprise five cases—all identical generators $\boldsymbol{S_4}$ (with only 1 $\boldsymbol{\omega}_\mathbf{g}$ and 24 symmetries described by $S_4$), three identical generators $\boldsymbol{K_{1,3}}$ (with 4 different $\boldsymbol{\omega}_\mathbf{g}$ and 6 symmetries described by $S_3$), two different pairs of identical generators $\boldsymbol{R_4}$ (with 6 different $\boldsymbol{\omega}_\mathbf{g}$ and 4 symmetries described by $V_4$), one pair of identical generators $\boldsymbol{P_4}$ (with 12 different $\boldsymbol{\omega}_\mathbf{g}$ and 2 symmetries described by $C_2$), and all different generators $\boldsymbol{X_4}$ (with 24 different $\boldsymbol{\omega}_\mathbf{g}$ and 1 symmetry described by $C_1$).

The relevant natural angular frequencies $\boldsymbol{\omega}_\mathbf{g}$, group order, and maximum angular frequency difference $||\boldsymbol{\omega}_\mathbf{g}||_\infty$ are detailed in Methods. Fig. 5 shows the Heatmaps of the vectors of natural angular frequencies $\boldsymbol{\omega}_\mathbf{g}$ for generators with different symmetries, including the cases of all identical generators $\boldsymbol{S_4}$, three identical generators $\boldsymbol{K_{1,3}}$, two different pairs of identical generators $\boldsymbol{R_4}$, one pair of identical generators $\boldsymbol{P_4}$, and all different generators $\boldsymbol{X_4}$. For a fair comparison, the maximum angular deviation sum is intentionally selected to be identical except for $\boldsymbol{S_4}$. The ranking list of components in terms of the maximum angular frequency difference satisfies $||\boldsymbol{\omega}_\mathbf{g_S4}||_\infty < ||\boldsymbol{\omega}_\mathbf{g_R4}||_\infty < ||\boldsymbol{\omega}_\mathbf{g_K13}||_\infty < ||\boldsymbol{\omega}_\mathbf{g_P4}||_\infty < ||\boldsymbol{\omega}_\mathbf{g_X4}||_\infty$, which almost reverses the ranking list of symmetries yet with an exception $\boldsymbol{R_4}$, which corresponds to the Klein Group and has a normal subgroup feature. Therefore, according to Eq. (4), the stability of power grids largely benefits from improved component symmetries. Note that the maximum angular frequency $||\boldsymbol{\omega}_\mathbf{g_R4}||_\infty$ is reduced due to the shared workload. Therefore, it is recommended to avoid the centralized effort of a single generator.

**(3) The Relationship between Symmetries and Stability**

As discussed, according to Eq. (4), from the perspective of network symmetries, the stability ranking list of network graphs in terms of the algebraic connectivity is $\boldsymbol{K_4} > \boldsymbol{C_4} > \boldsymbol{K_{1,3}} > \boldsymbol{R_4} > \boldsymbol{P_4} > \boldsymbol{X_4}$, which confirms that stability agrees with symmetries. From the perspective of component symmetries, the stability ranking list of components in terms of the maximum angular frequency follows $\boldsymbol{S_4} > \boldsymbol{R_4} > \boldsymbol{K_{1,3}} > \boldsymbol{P_4} > \boldsymbol{X_4}$. In this case, the stability ranking list roughly agrees with the symmetry ranking list except for $\boldsymbol{R_4}$, which could be contributed by its special position as a normal subgroup among other factors.

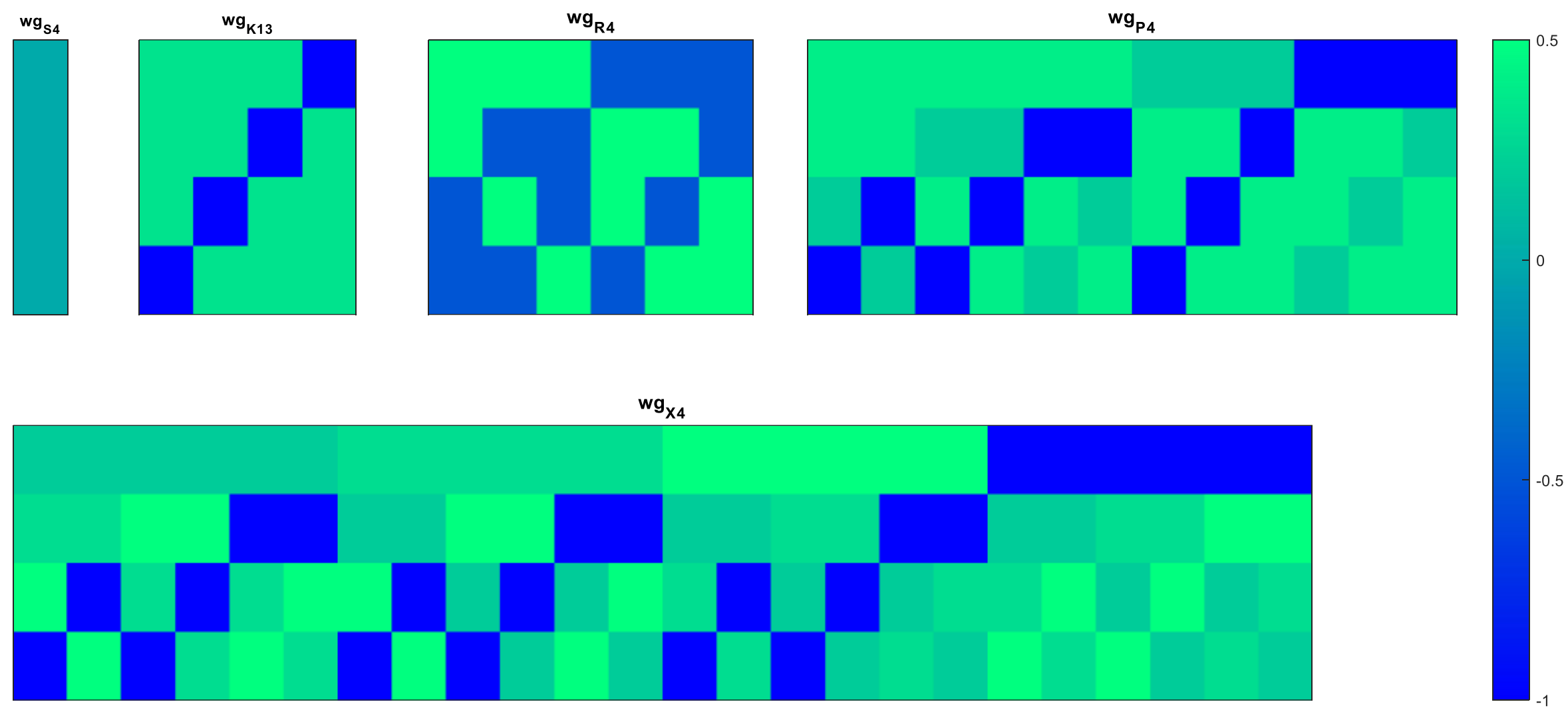

**Fig. 5 Heatmaps of the vectors of natural angular frequencies $\omega_g$ for generators with different symmetries, including the cases of all identical generators $S_4$, three identical generators $K_{1,3}$, two different pairs of identical generators $R_4$, one pair of identical generators $P_4$, and all different generators $X_4$.** For a fair comparison, the maximum deviation sum is intentionally selected to be identical except for $S_4$. $S_4$ features twenty-four symmetries and unique $\omega_g$; $K_{1,3}$ features six symmetries and four different $\omega_g$; $R_4$ features four symmetries and six different $\omega_g$; $P_4$ features two symmetries and twelve different $\omega_g$; $X_4$ features one symmetry and twenty-four different $\omega_g$. The ranking list of components in terms of the maximum angular frequency difference satisfies $||\omega_{g_S4}||_\infty < ||\omega_{g_R4}||_\infty < ||\omega_{g_K13}||_\infty < ||\omega_{g_P4}||_\infty < ||\omega_{g_X4}||_\infty$, which almost reverses the ranking list of symmetries yet with an exception $R_4$, which corresponds to the Klein Group and has a normal subgroup feature.

For better validation and visualization, according to Eq. (3), we further derive the stability evaluation terms $\mathbf{L}^{\dagger}\boldsymbol{\omega}_{\mathbf{g}}$ of all the combinations of different networks and components. In each case, $\left\|\mathbf{L}^{\dagger}\boldsymbol{\omega}_{\mathbf{g}}\right\|_{\mathbf{E},\infty}$ may not be unique, where the detailed results are provided in Methods.

Fig. 6 illustrates the heatmaps of the key stability evaluation term $\mathbf{L}^{\dagger}\boldsymbol{\omega}_{\mathbf{g}}$ of the power grids with different combinations of network and component symmetries. As shown, the figures in each column differ by network graphs, including the complete graph $K_4$, cyclic graph $C_4$, star graph $K_{1,3}$, rectangular graph $R_4$, path graph $P_4$, and random graph $X_4$. The figures in each row differ by components, including the cases of all identical generators $S_4$, three identical generators $K_{1,3}$, two different pairs of identical generators $R_4$, one pair of identical generators $P_4$, and all different generators $X_4$. The stability of a power grid is encoded by the colour contrast of heatmaps, where the larger the colour contrast, the higher chance of instability.

For better comparison, we tabulate the results of stability indices $\left\|\mathbf{L}^{\dagger}\boldsymbol{\omega}_{\mathbf{g}}\right\|_{\mathbf{E},\infty,\mathrm{avg}}$ and $\left\|\mathbf{L}^{\dagger}\boldsymbol{\omega}_{\mathbf{g}}\right\|_{\mathbf{E},\infty,\max}$ (in the bracket) in Table II. Note that the grid stability is reversely proportional to the value of stability indices listed in Table II. From the perspective of network symmetries, the ranking list of power grids in terms of stability should be $K_4 > C_4 > K_{13} > R_4 > P_4 > X_4$, which agrees with the previous results. From the perspective of component symmetries, the ranking list of power grids in terms of stability should be $S_4 > R_4 > K_{13} > P_4 > X_4$. Once again, the stability ranking list roughly agrees with the symmetry ranking list except for $R_4$, which exhibits slightly better stability than $K_{1,3}$ in most cases, but reverse conditions do exist.

## V. Simulation and Experimental Verifications

In this part, we will validate the analysis on the stability of power grids in terms of network symmetries and component symmetries with the parameters listed in Table III sequentially via simulations and experimentals. To yield fair comparisons, line impedances are unified by multiplication of the line inductance reference by the relevant coefficient.

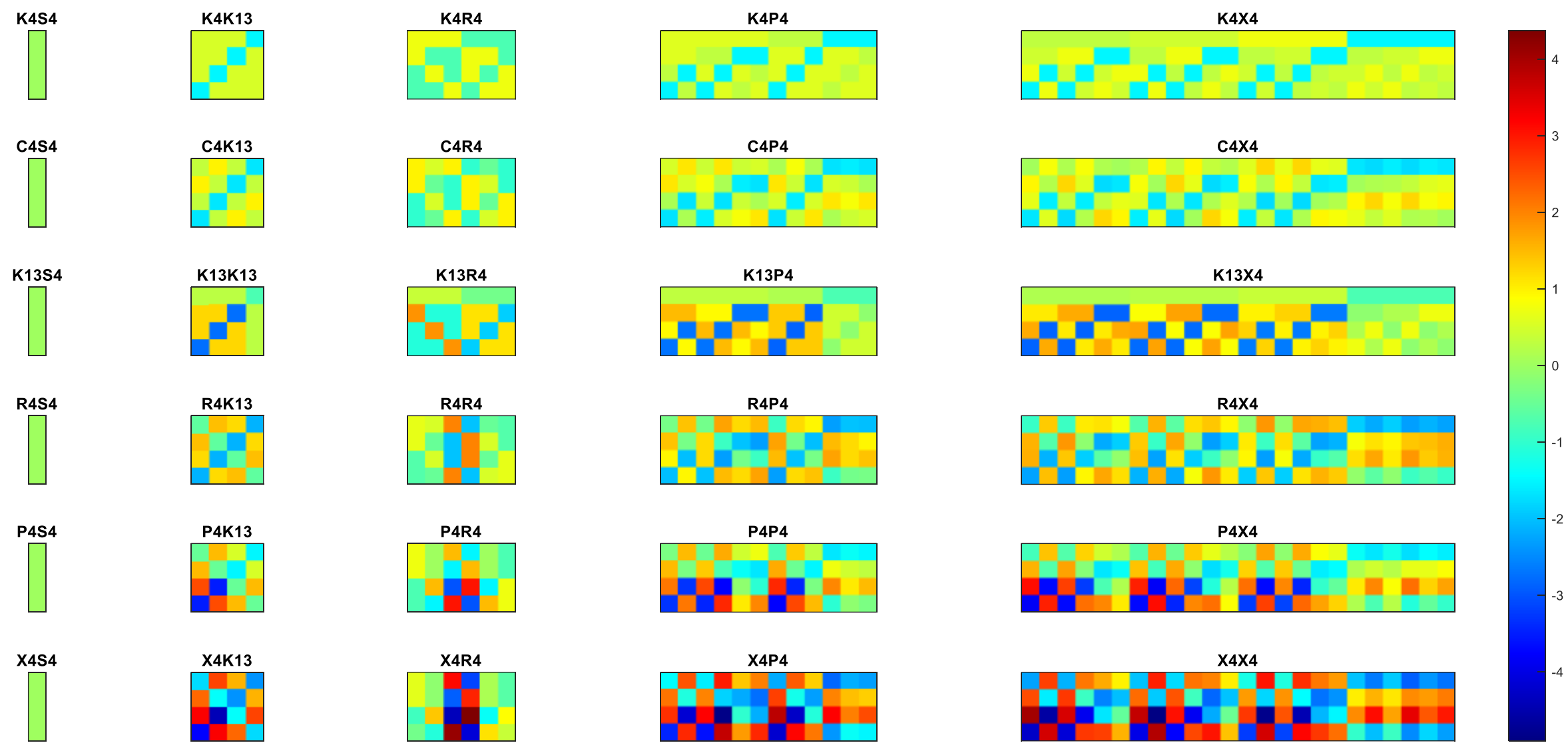

**Fig. 6 Heatmaps of the key stability evaluation terms $\mathbf{L}_i\omega_g$ of the power grids with different combinations of network and component symmetries.** The figures in each column differ by network graphs, including the complete graph $\boldsymbol{K_4}$, cyclic graph $\boldsymbol{C_4}$, star graph $\boldsymbol{K_{1,3}}$, rectangular graph $\boldsymbol{R_4}$, path graph $\boldsymbol{P_4}$, and random graph $\boldsymbol{X_4}$. The figures in each row differ by components, including the cases of all identical generators $\boldsymbol{S_4}$, three identical generators $\boldsymbol{K_{1,3}}$, two different pairs of identical generators $\boldsymbol{R_4}$, one pair of identical generators $\boldsymbol{P_4}$, and all different generators $\boldsymbol{X_4}$. The stability of power networks is encoded by the colour contrast of heatmaps, where the larger the colour contrast, the higher chance of instability. From the perspective of network symmetries, the ranking list of power grids in terms of stability should be $\boldsymbol{K_4} > \boldsymbol{C_4} > \boldsymbol{K_{13}} > \boldsymbol{R_4} > \boldsymbol{P_4} > \boldsymbol{X_4}$. From the perspective of component symmetries, the ranking list of power grids in terms of stability should be $\boldsymbol{S_4} > \boldsymbol{R_4} > \boldsymbol{K_{13}} > \boldsymbol{P_4} > \boldsymbol{X_4}$. Notably, the stability ranking list roughly agrees with the symmetry ranking list except for $\boldsymbol{R_4}$.

**Table II Results of the average and maximum (in the bracket) stability indices, with smaller values representing better stability.** The first column refers to graphs with different network symmetries. The first row refers to generators with different component symmetries**.** The remaining elements correspond to the relevant results of the average $\left\|\mathbf{L}^{\dagger}\boldsymbol{\omega}_{\mathbf{g}}\right\|_{\mathbf{E},\infty,\text{avg}}$ and maximum $\left\|\mathbf{L}^{\dagger}\boldsymbol{\omega}_{\mathbf{g}}\right\|_{\mathbf{E},\infty,\max}$ (in the bracket) stability indices.

| Components / Graphs | $S_4$ | $K_{1,3}$ | $R_4$ | $P_4$ | $X_4$ |
|---|---|---|---|---|---|
| $K_4$ | 0 (0) | 2 (2) | 1.5 (1.5) | 2.1 (2.1) | 2.25 (2.25) |
| $C_4$ | 0 (0) | 2.667 (2.667) | 1.667 (2) | 2.667 (2.8) | 2.667 (3) |
| $K_{1,3}$ | 0 (0) | 3.25 (4) | 3 (3) | 3.45 (4.2) | 3.60 (4.5) |
| $R_4$ | 0 (0) | 3.556 (3.556) | 2.111 (4) | 3.556 (4) | 3.583 (4.133) |
| $P_4$ | 0 (0) | 4.5 (6) | 3.5 (6) | 4.5 (6.6) | 4.5 (6.9) |
| $X_4$ | 0 (0) | 6.0833 (7) | 4.1667 (8.5) | 6.0833 (8.6) | 6.0833 (8.8) |

## (1) Simulation Verifications

First, we conducted simulations for verification. Specifically, we select a case of generators with three identical generators (i.e., $\boldsymbol{K_{13}}$) and evaluate the stability of power grids from the perspective of network symmetries. For visualization, Fig. 7 presents the simulated waveforms of the phase angles $\theta_{g1}$, $\theta_{g2}$, $\theta_{g3}$, and $\theta_{g4}$ as well as the active power outputs $p_{g1}$, $p_{g2}$, $p_{g3}$, and $p_{g4}$ of individual generators, where the solid line "—", the dash line "– –", the dotted line "- -", and the dash dotted line "– -" refer to the first, second, third, and fourth generators, respectively. In simulations, generators start to operate at $t = 0.2$ s and change their active power references at $t = 5$ s, where $p_{g_ref}$ stands for the total power deviation. In the case of three identical generators, $p_{g_ref} = 3p_{g1_ref} = 3p_{g2_ref} = 3p_{g3_ref} = -p_{g4_ref}$. Notably, the increase of $p_{g_ref}$ boosts the power deviation and hence the chance of instability. As shown, generators can run stably before $t = 5$ s and become unstable afterwards. Clearly, the networks with different symmetries feature different stability, as indicated by the boundary values of $p_{g_ref}$. For the network $\boldsymbol{K_4}$, it performs the best in terms of stability and maintains stable when $p_{g_ref} \leq 5610$ W. Next comes the network $\boldsymbol{C_4}$, which ranks the second with a boundary $p_{g_ref}$ of 5540 W. In this specific case, the network $\boldsymbol{K_{1,3}}$ with $p_{g_ref} = 4480$ W outweighs $\boldsymbol{R_4}$ with $p_{g_ref} = 3930$ W. Both $\boldsymbol{R_4}$ and $\boldsymbol{K_{1,3}}$ possess better stability than $\boldsymbol{P_4}$, whose corresponding stable region is $p_{g_ref} \leq 3700$ W. Finally, the network $\boldsymbol{X_4}$ suffers from poor stability with $p_{g_ref} = 2730$ W. By taking a closer look at the phase angle curves of individual networks, we notice that generators in general form two groups of divergent phase angles after $p_{g_ref}$ exceeds a certain limitation, thereby implying the loss of synchronism.

Table III System parameters and control parameters of generators in simulations and experiments.

| Variables | Symbols | Values |
| --- | --- | --- |
| Number of Generators | $n$ | 4 |
| Rated Frequency | $f_{g0}$ | 50 Hz |
| Rated Angular Frequency | $\omega_{g0}$ | 314 rad/s |
| Line Inductance Reference | $L_{g_ref}$ | 10 mH |
| Line Resistance | $R_g$ | 1 Ω |
| Reference Power | $p_{g_ref}$ | Case Dependent |
| Rated Power | $S_g$ | 1000 VA |
| Damping Coefficient | $D_g$ | 100 |
| Terminal Voltage Amplitude | $v_g$ | 110 Vrms |

Fig. 8 demonstrates the simulated waveforms of the active power outputs $p_{g1}$, $p_{g2}$, and $p_{g3}$ with a small change of the total power deviation $p_{g_ref}$ at $t = 0.2$ s in power networks with different network symmetries, showing their recovery speed in terms of the settling time $t_{set}$ under such small disturbances. As shown, $\boldsymbol{K_4}$ features a settling time of $t_{set} = 0.3$ s, $\boldsymbol{C_4}$ features a settling time of $t_{set} = 0.4$ s, $\boldsymbol{K_{1,3}}$ features a settling time of $t_{set} = 0.45$ s, $\boldsymbol{R_4}$ features a settling time of $t_{set} = 0.7$ s, $\boldsymbol{P_4}$ features a settling time of $t_{set} = 0.8$ s, and $\boldsymbol{X_4}$ features a settling time of $t_{set} = 1.6$ s. The ranking list in dynamic performances agrees well with that in terms of network symmetries, which verifies the effectiveness of the analysis with algebraic connectivity $\lambda_2$.

The complete graph $\boldsymbol{K_4}$ serves as an example to evaluate the stability of power grids from the perspective of component symmetries. As shown by the simulation waveforms in Fig. 9, generators start to operate at $t = 0.2$ s and change their active power references at $t = 5$ s, after which the divergence of phase angles can be noticed. Once again, $p_{g_ref}$ stands for the total power deviation, which remains unchanged in all cases except for $\boldsymbol{S_4}$. In comparison, the case of all identical generators $\boldsymbol{S_4}$ enjoys the best stability, as the power grid is always sable regardless of $p_{g_ref}$. The case of two different pairs of identical generators $\boldsymbol{R_4}$ (whose stable region is $p_{g_ref} \leq 7690$ W) features better stability than the case of three identical generators $\boldsymbol{K_{1,3}}$ (whose stable region is $p_{g_ref} \leq 5610$ W). $\boldsymbol{P_4}$ and $\boldsymbol{X_4}$ feature a stable region of $p_{g_ref} \leq 5590$ W and $p_{g_ref} \leq 5580$ W, respectively, which agree well with theoretical analysis.

**(2) Experimental Verifications**

Next, we conducted experiments for further verifications, while testing parameters and conditions remain unchanged as those in simulations. Fig. 10 visualizes the experimental setup, where a hardware-in-loop testing platform from EasyGo implements the power grid and its control algorithm, while an eight-channel oscilloscope from LeCroy captured all the experimental waveforms. Figs. 11–13 demonstrate the experimental results for verification of the relationship between the network symmetries and power grid stability, the network symmetries and system dynamics, and the component symmetries and power grid stability, respectively, which are highly consistent with their corresponding simulations in Figs. 7–9.

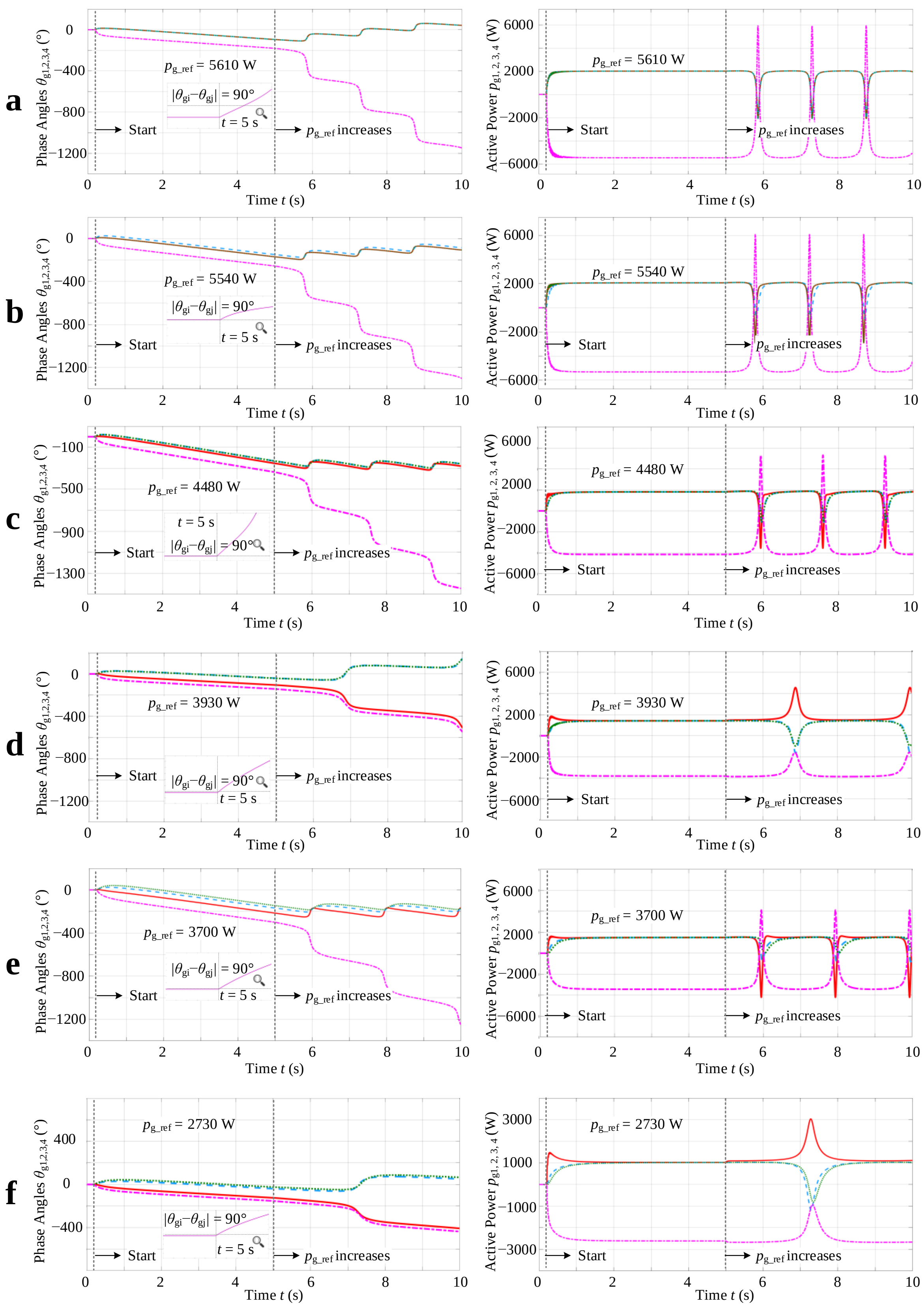

**Fig. 7 Simulated waveforms of the phase angles $\theta_{g1}$, $\theta_{g2}$, $\theta_{g3}$, $\theta_{g4}$ and active power outputs $p_{g1}$, $p_{g2}$, $p_{g3}$, and $p_{g4}$ in the case of three identical generators $K_{13}$ with the change of the total power deviation $p_{g_ref} = 3p_{g1_ref} = 3p_{g2_ref} = 3p_{g3_ref} = -p_{g4_ref}$ at $t$ = 5 s in power networks with different network symmetries, including the complete graph $K_4$, cyclic graph $C_4$, star graph $K_{1,3}$, rectangular graph $R_4$, path graph $P_4$, and random graph $X_4$.** **a** $K_4$ features a stable region of $p_{g_ref} \leq 5610$ W. **b** $C_4$ features a stable region of $p_{g_ref} \leq 5540$ W. **c** $K_{1,3}$ features a stable region of $p_{g_ref} \leq 4480$ W. **d** $R_4$ features a stable region of $p_{g_ref} \leq 3930$ W. **e** $P_4$ features a stable region of $p_{g_ref} \leq 3700$ W. **f** $X_4$ features a stable region of $p_{g_ref} \leq 2730$ W.

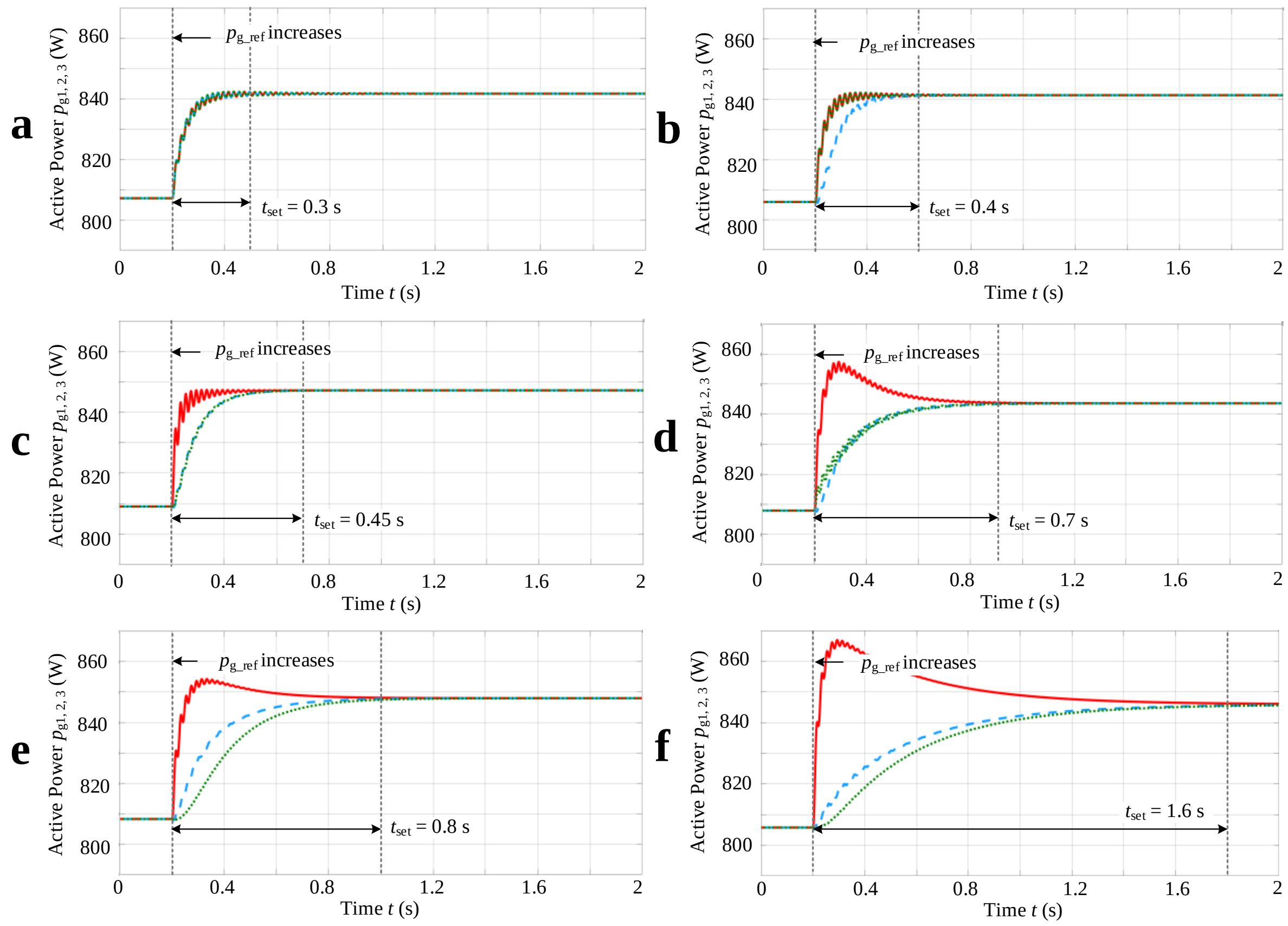

**Fig. 8 Simulated waveforms of the active power outputs $p_{g1}$, $p_{g2}$, and $p_{g3}$ in the case of three identical generators $K_{13}$ with a small change of the total power deviation $p_{g_ref}$ at $t$ = 0.2 s in power networks with different network symmetries, including the complete graph $K_4$, cyclic graph $C_4$, star graph $K_{1,3}$, rectangular graph $R_4$, path graph $P_4$, and random graph $X_4$, showing their recovery speed in terms of the settling time $t_{set}$ under such small disturbances.** **a** $K_4$ features a settling time of $t_{set}$ = 0.3 s. **b** $C_4$ features a settling time of $t_{set}$ = 0.4 s. **c** $K_{1,3}$ features a settling time of $t_{set}$ = 0.45 s. **d** $R_4$ features a settling time of $t_{set}$ = 0.7 s. **e** $P_4$ features a settling time of $t_{set}$ = 0.8 s. **f** $X_4$ features a settling time of $t_{set}$ = 1.6 s.

Both simulated and experimental results verify the effectiveness of theoretical analyses.

## Discussions

The stability of power grids generally depends on the synchronization of generators, which is subjected to a conflict between the coupling strength and the misalignment of power references. Once the coupling strength outweighs power reference differences, generators tend to synchronize, leading to stabilized power grids, and vice versa. This article investigates how grid symmetries affect the synchronized conditions of generators and hence the stability of power grids by innovatively using group theory as combined with graph theory. It is revealed that grid symmetries comprise two aspects—network symmetries and component symmetries. We quantitatively investigate network symmetries via the automorphism groups of network graphs, where the order of automorphism groups is a major indicator of symmetries. In parallel, we use the subgroups of symmetric groups to investigate component symmetries. In the perspective of network symmetries, where grid stability and synchronization speed are dictated by the algebraic connectivity, the stability ranking list of grids with four generators is $\boldsymbol{K_4} > \boldsymbol{C_4} > \boldsymbol{K_{1,3}} > \boldsymbol{R_4} > \boldsymbol{P_4} > \boldsymbol{X_4}$, which fully agrees with the list of network symmetries. In the perspective of component symmetries, it is found that improved component symmetries largely translate into better system stability yet with $\boldsymbol{R_4}$ as an exceptional case, which corresponds to the Klein Group and has a normal subgroup feature, where it is recommended to avoid the centralized effort of a single generator. Moreover, we disclose that factors other than symmetries also impact grid stability, e.g., coupling strength, power imbalance, edge numbers, and node degrees, which agree well with previous findings. These conclusions are verified by simulation results under large disturbances, which have significant implications for the control of modern power grids with distributed resources and controllable generators, including energy storage and renewable power generators interfaced by grid-following/-forming converters.

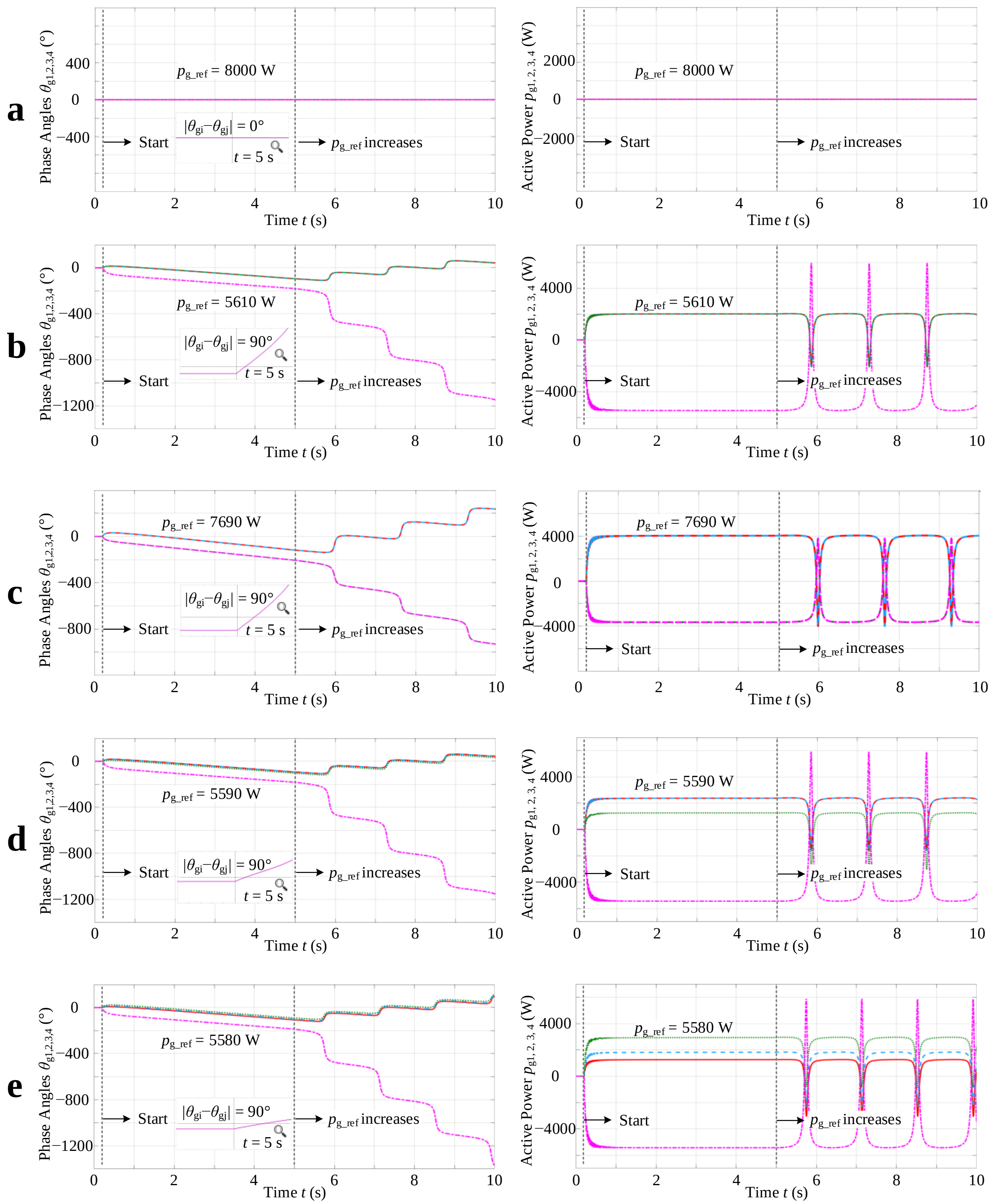

**Fig. 9 Simulated waveforms of the phase angles $\theta_{g1}$, $\theta_{g2}$, $\theta_{g3}$, $\theta_{g4}$ and active power outputs $p_{g1}$, $p_{g2}$, $p_{g3}$, and $p_{g4}$ for the complete graph $K_4$ with the change of the total power deviation $p_{g_ref}$ at $t$ = 5 s in power networks with different component symmetries, including all identical generators $S_4$, three identical generators $K_{1,3}$, two different pairs of identical generators $R_4$, one pair of identical generators $P_4$, and all different generators $X_4$. a** $S_4$ is always stable. **b** $K_{1,3}$ features a stable region of $p_{g_ref} \le 5610$ W. **c** $R_4$ features a stable region of $p_{g_ref} \le 7690$ W. **d** $P_4$ features a stable region of $p_{g_ref} \le 5590$ W. **e** $X_4$ features a stable region of $p_{g_ref} \le 5580$ W.

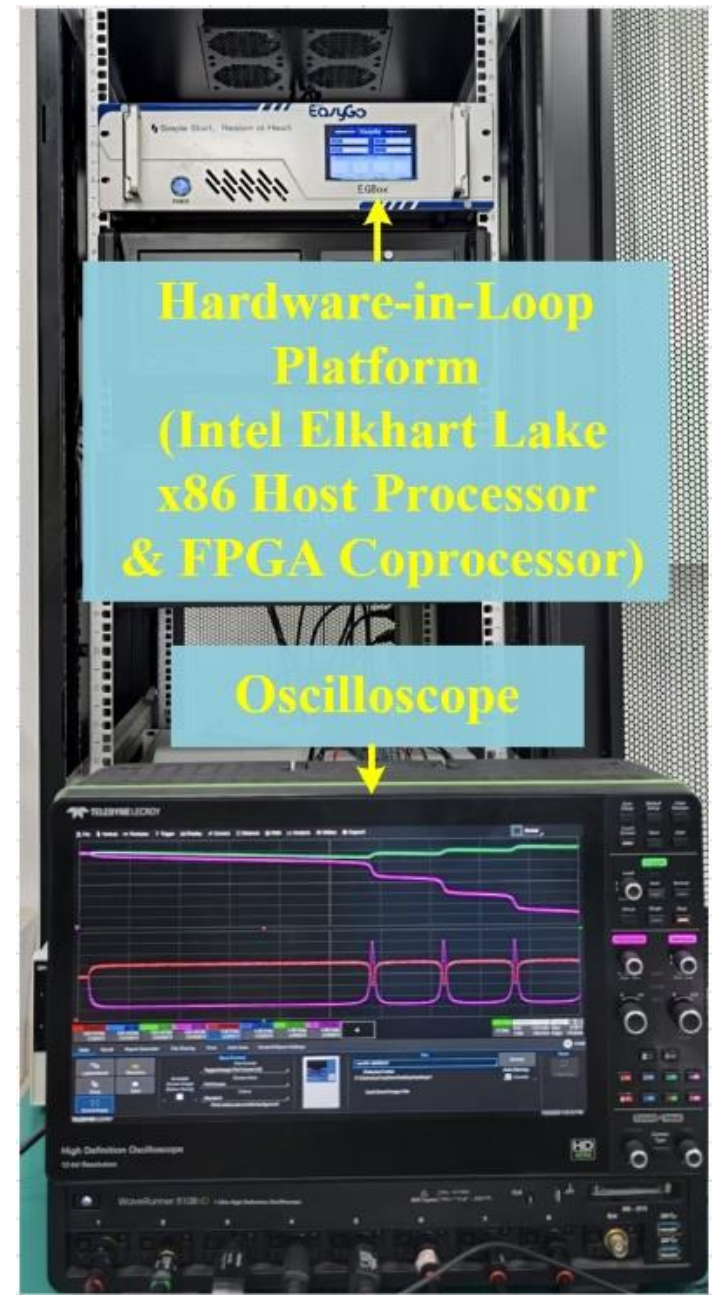

**Fig. 10 A photo of the experimental setup.** A hardware-in-loop testing platform implements the power grid and its control algorithm, while an eight-channel oscilloscope captured all the experimental waveforms.

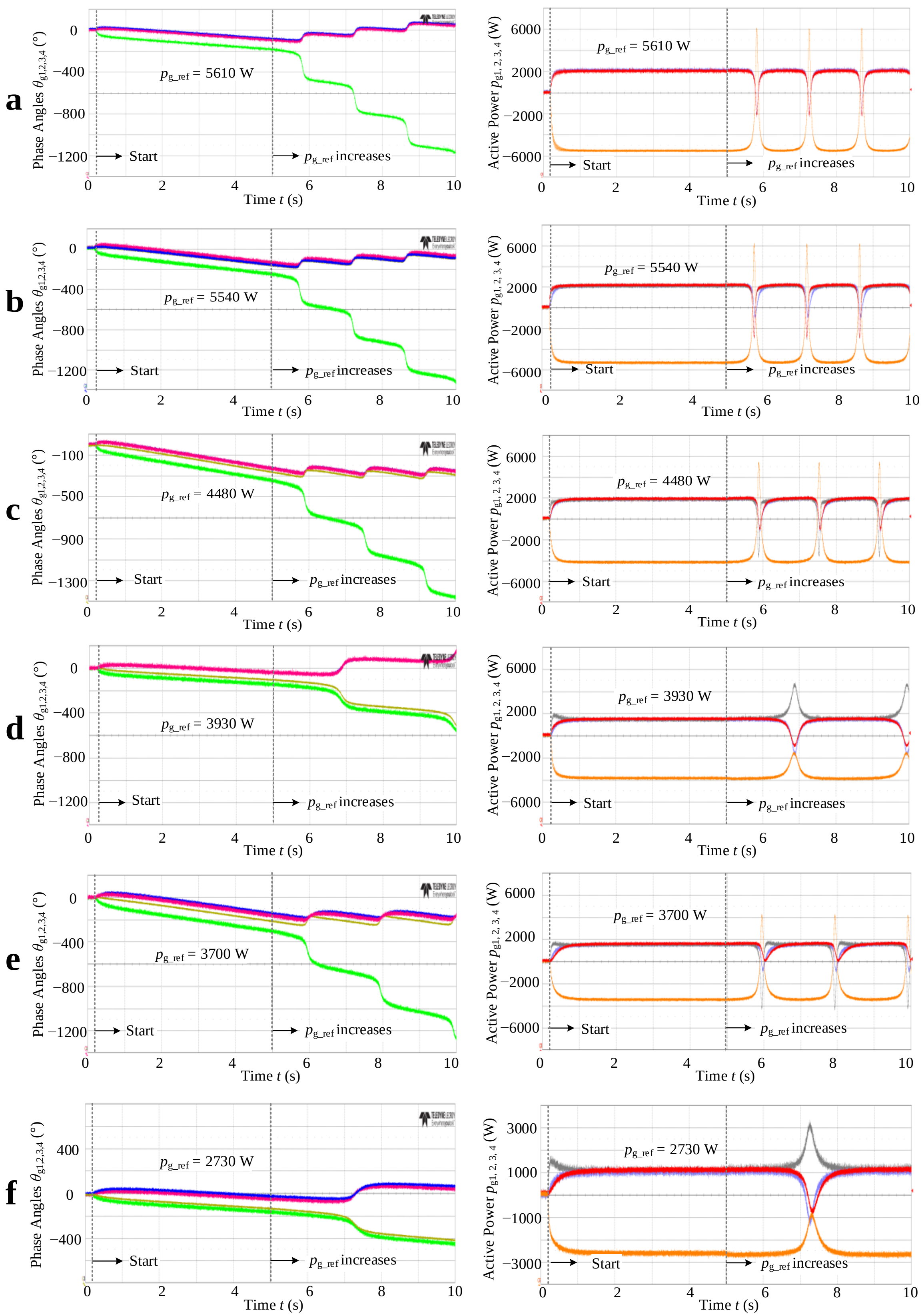

**Fig. 11 Experimental waveforms of the phase angles $\theta_{g1}$, $\theta_{g2}$, $\theta_{g3}$, $\theta_{g4}$ and active power outputs $p_{g1}$, $p_{g2}$, $p_{g3}$, and $p_{g4}$ in the case of three identical generators $K_{13}$ with the change of the total power deviation $p_{g_ref} = 3p_{g1_ref} = 3p_{g2_ref} = 3p_{g3_ref} = -p_{g4_ref}$ at $t = 5$ s in power networks with different network symmetries, including the complete graph $K_4$, cyclic graph $C_4$, star graph $K_{1,3}$, rectangular graph $R_4$, path graph $P_4$, and random graph $X_4$.** **a** $K_4$ features a stable region of $p_{g_ref} \leq 5610$ W. **b** $C_4$ features a stable region of $p_{g_ref} \leq 5540$ W. **c** $K_{1,3}$ features a stable region of $p_{g_ref} \leq 4480$ W. **d** $R_4$ features a stable region of $p_{g_ref} \leq 3930$ W. **e** $P_4$ features a stable region of $p_{g_ref} \leq 3700$ W. **f** $X_4$ features a stable region of $p_{g_ref} \leq 2730$ W.

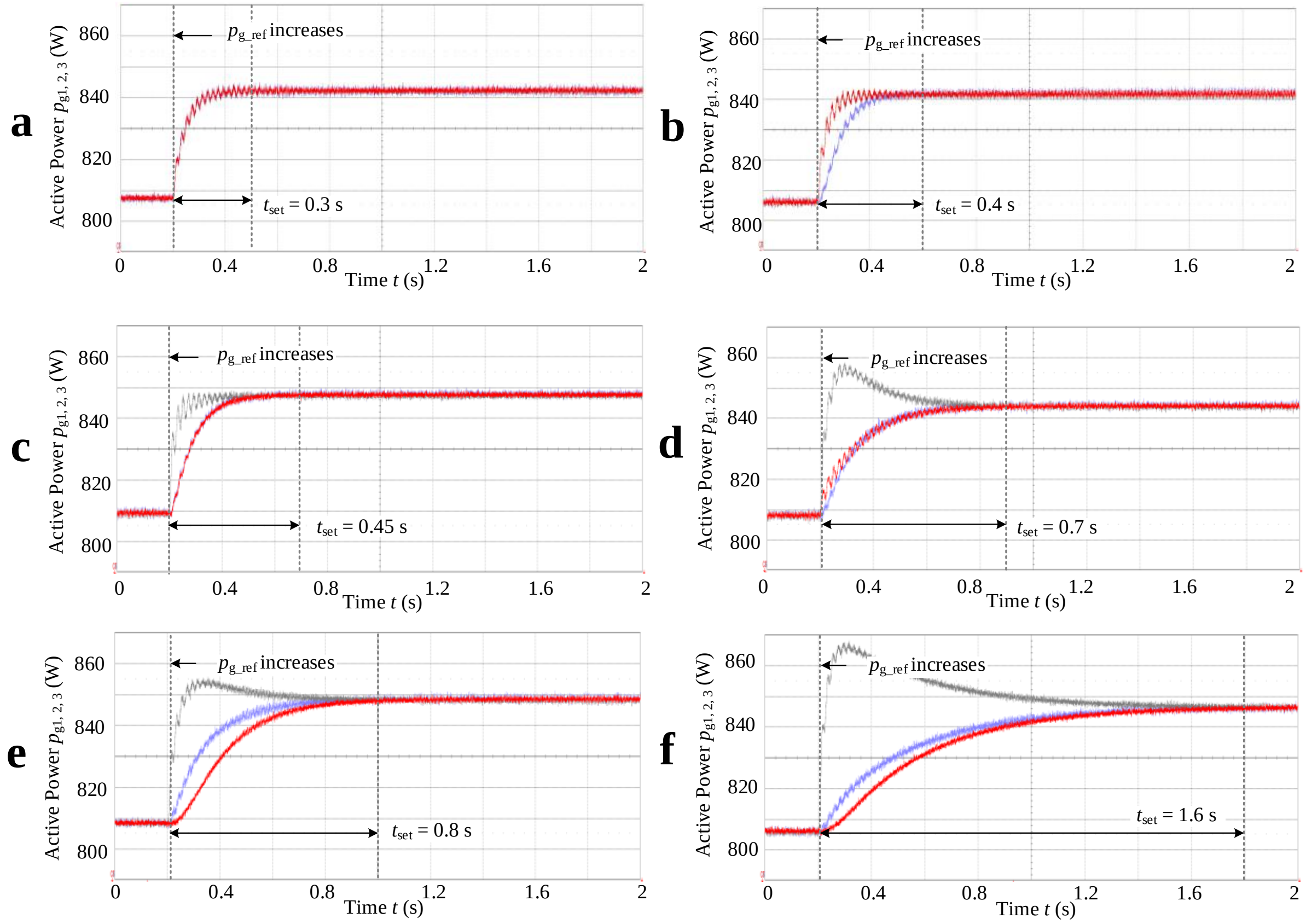

**Fig. 12 Experimental waveforms of the active power outputs $p_{g1}$, $p_{g2}$, and $p_{g3}$ in the case of three identical generators $K_{13}$ with a small change of the total power deviation $p_{g_ref}$ at $t$ = 0.2 s in power networks with different network symmetries, including the complete graph $K_4$, cyclic graph $C_4$, star graph $K_{1,3}$, rectangular graph $R_4$, path graph $P_4$, and random graph $X_4$, showing their recovery speed in terms of the settling time $t_{set}$ under such small disturbances.** **a** $K_4$ features a settling time of $t_{set}$ = 0.3 s. **b** $C_4$ features a settling time of $t_{set}$ = 0.4 s. **c** $K_{1,3}$ features a settling time of $t_{set}$ = 0.45 s. **d** $R_4$ features a settling time of $t_{set}$ = 0.7 s. **e** $P_4$ features a settling time of $t_{set}$ = 0.8 s. **f** $X_4$ features a settling time of $t_{set}$ = 1.6 s.

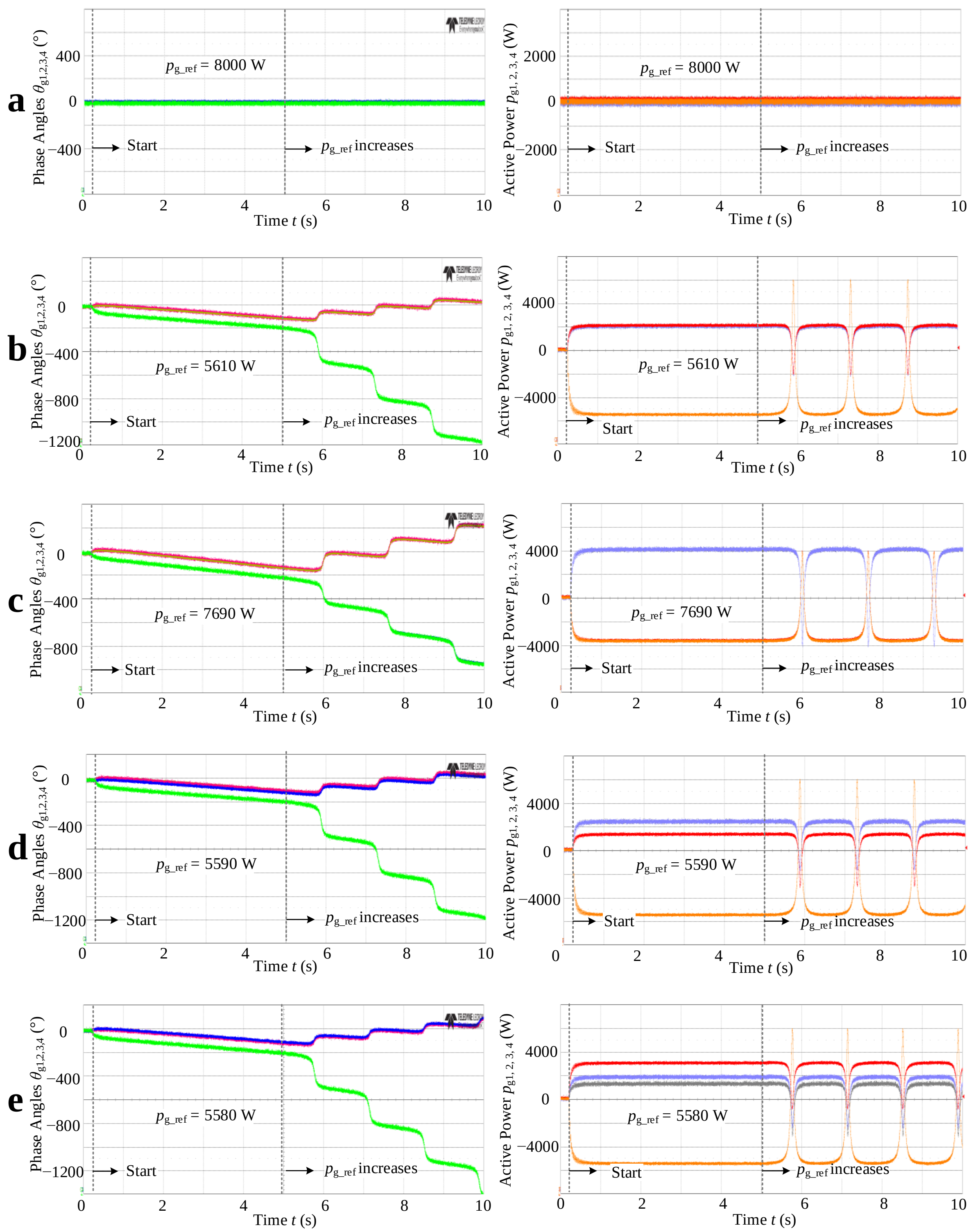

**Fig. 13 Experimental waveforms of the phase angles $\theta_{g1}$, $\theta_{g2}$, $\theta_{g3}$, $\theta_{g4}$ and active power outputs $p_{g1}$, $p_{g2}$, $p_{g3}$, and $p_{g4}$ for the complete graph $K_4$ with the change of the total power deviation $p_{g_ref}$ at $t$ = 5 s in power networks with different component symmetries, including all identical generators $S_4$, three identical generators $K_{1,3}$, two different pairs of identical generators $R_4$, one pair of identical generators $P_4$, and all different generators $X_4$. a** $S_4$ is always stable. **b** $K_{1,3}$ features a stable region of $p_{g_ref} \leq 5610$ W. **c** $R_4$ features a stable region of $p_{g_ref} \leq$ 7690 W. **d** $P_4$ features a stable region of $p_{g_ref} \leq 5590$ W. **e** $X_4$ features a stable region of $p_{g_ref} \leq 5580$ W.